\documentclass[12pt]{article}
\pdfoutput=1
\usepackage{color}
\usepackage{amssymb,amsmath,bm}
\usepackage{epsf}
\usepackage{epsfig}
\usepackage{afterpage}
\usepackage{longtable}
\usepackage{cite}
\usepackage{latexsym,mathrsfs}
\usepackage{graphics}
\usepackage{url}
\usepackage{paralist}
\usepackage{subcaption}

\usepackage{float}

\newcommand{\B}{{Z'}}
\usepackage{booktabs}
\definecolor{red}{cmyk}{0,1,1,0.4}
\definecolor{darkgreen}{rgb}{0.0,0.6,0.0}
\definecolor{cDarkGrey}{RGB}{91,91,91}
\definecolor{cGrey}{RGB}{245,243,238}
\definecolor{cBlue}{RGB}{0,110,191}
\definecolor{cLightBlue}{RGB}{214,237,252}
\definecolor{cRed}{RGB}{196,0,100}
\definecolor{cLightRed}{RGB}{254,222,237}
\definecolor{cGreen}{RGB}{0,166,80}
\definecolor{cLightGreen}{RGB}{254,222,237}
\definecolor{cOrange}{RGB}{221,74,44}
\definecolor{cLightOrange}{RGB}{255,215,210}
\definecolor{cPurple}{RGB}{93,35,125}
\definecolor{cLightPurple}{RGB}{241,230,252}
\definecolor{cYellow}{RGB}{252,191,10}
\definecolor{cISSRBlue}{RGB}{0,111,174}
\definecolor{cISSRGrey}{RGB}{167,169,172}

\newcommand{\IM}{\text{Im}}

\newcommand{\imlt}{\IM\lambda_t}

\newcommand{\tev}{\, {\rm TeV}}
\newcommand{\gev}{\, {\rm GeV}}
\newcommand{\mev}{\, {\rm MeV}}
\def\B{\mathcal{B}}
\newcommand{\vcb}{|V_{cb}|}

\newcommand{\vub}{|V_{ub}|}

\newcommand{\vus}{|V_{us}|}
\def \refeq#1{(\ref{#1})}
\def\ksm{K_S\to\mu\bar\mu}
\def\klm{K_L\to\mu\bar\mu}

\newcommand{\kepe}{\kappa_{\varepsilon^\prime}}
\newcommand{\keps}{\kappa_{\varepsilon}}
\def\klpll{K_L\to\pi^0\ell\bar\ell}
\def\epe{\varepsilon'/\varepsilon}
\newcommand{\beq}{\begin{equation}}
\newcommand{\eeq}{\end{equation}}
\newcommand{\be}{\begin{equation}}
\newcommand{\ee}{\end{equation}}
\newcommand{\bi}{\begin{itemize}}
\newcommand{\ei}{\end{itemize}}
\newcommand{\ba}{\begin{array}}
\newcommand{\ea}{\end{array}}
\newcommand{\beqa}{\begin{eqnarray}}
\newcommand{\eeqa}{\end{eqnarray}}
\newcommand{\bea}{\begin{eqnarray}}
\newcommand{\eea}{\end{eqnarray}}
\newcommand{\beqn}{\begin{eqnarray}}
\newcommand{\eeqn}{\end{eqnarray}}

\newcommand{\eps}{\epsilon}

\newcommand{\muEW}{{\mu_{\rm EW}}}

\newcommand{\BR}{{\cal B}}

\newcommand{\epsK}{\varepsilon_K}

\definecolor{red}{cmyk}{0,1,1,0.4}

\def\kpn{K^+\rightarrow\pi^+\nu\bar\nu}
\def\klpn{K_{L}\rightarrow\pi^0\nu\bar\nu}

\usepackage{fancyhdr}
\advance \headheight by 3.0truept       

\begin{document}

\begin{flushright}
    {AJB-22-4}\\

\end{flushright}

\medskip

\begin{center}
{\LARGE\bf
\boldmath{On the Importance of  Rare Kaon Decays:\\
A Snowmass 2021 White Paper}}
\\[0.8 cm]
{\bf Jason~Aebischer$^{a}$,  Andrzej~J.~Buras$^{b}$ and
Jacky Kumar$^{b}$
 \\[0.5 cm]}
{\small
$^a$ Physik-Institut, Universit\"at Z\"urich, CH-8057 Z\"urich, Switzerland\\
$^b$TUM Institute for Advanced Study, Lichtenbergstr. 2a, D-85747 Garching, Germany\\}
\end{center}

\vskip0.41cm

\noindent
\begin{abstract}
  We stress the importance of  precise measurements of
  rare decays $K^+\rightarrow\pi^+\nu\bar\nu$, $K_L\rightarrow\pi^0\nu\bar\nu$, $K_{L,S}\to\mu^+\mu^-$ and  $K_{L,S}\to\pi^0\ell^+\ell^-$ for   the search of new physics (NP). This includes both branching ratios and the distributions in $q^2$, the invariant mass-squared of the neutrino system in the case of $K^+\rightarrow\pi^+\nu\bar\nu$ and $K_L\rightarrow\pi^0\nu\bar\nu$ and of the $\ell^+\ell^-$ system
in the case of the remaining decays.
  In particular the correlations
  between these  observables and their correlations
  with the ratio $\varepsilon'/\varepsilon$ in $K_L\to\pi\pi$ decays, the CP-violating
  parameter $\varepsilon_K$ and the $K^0-\bar K^0$ mass difference $\Delta M_K$,
  should help to disentangle the nature of possible NP. We stress the
  strong sensitivity of all observables with the exception of $\Delta M_K$
  to the CKM parameter  $|V_{cb}|$ and list a number of $|V_{cb}|$-independent ratios within the SM which exhibit rather different dependences on the angles $\beta$ and
  $\gamma$ of the unitarity triangle. The particular role of these decays
  in probing very short distance scales far beyond the ones explored at the LHC
  is emphasized. In this context the role of the Standard Model Effective Field Theory (SMEFT) is very important.
  We also address briefly the issue of the footprints of   Majorana neutrinos in  $K^+\rightarrow\pi^+\nu\bar\nu$ and $K_L\rightarrow\pi^0\nu\bar\nu$.

\end{abstract}

\thispagestyle{empty}
\newpage
\tableofcontents
\newpage
\setcounter{page}{1}

\section{Introduction}
Rare decays of Kaons played already for decades a very important role in testing the Standard Model (SM) and in the search for new physics (NP). In this decade significant progress on these decays will be made, in particular through
experiments at CERN (NA62, LHCb) \cite{Ceccucci:2018bnw,Cerri:2018ypt,Bediaga:2018lhg},  J-PARC (KOTO) \cite{Ahn:2018mvc} and later also KLEVER \cite{Ambrosino:2019qvz} at CERN.
Among the rare Kaon decays considered by us
$\kpn$, $K_L\to\mu^+\mu^-$ and $K_S\to\pi^0\ell^+\ell^-$
are CP conserving while $\klpn$,  $K_S\to\mu^+\mu^-$ and  $K_L\to\pi^0\ell^+\ell^-$ proceed in the SM and in many of the beyond SM (BSM) scenarios governed
by vector and axial-vector currents only in the presence of CP violation. The latter fact makes
the search for these decays very important with the goal to find new sources of
CP violation possibly responsible for the matter-antimatter asymmetry in the
universe. A recent extensive review of these decays can be found in \cite{Buras:2020xsm}. However, a less known fact should be emphasized here. In the presence of scalar currents $\klpn$,  $K_S\to\mu^+\mu^-$ and  $K_L\to\pi^0\ell^+\ell^-$ can proceed also without any sources of CP violation \cite{Kiyo:1998zm}.

Within the SM these decays are loop-induced
semileptonic FCNC processes, receiving  only contributions from
$Z^0$-penguin and box diagrams, in particular with $W^\pm$ and top quark exchanges.
A  very important virtue of $\kpn$ and $\klpn$ decays
is their clean theoretical character.
This is related to the fact that
the low energy hadronic
matrix elements required for the calculations of their branching ratios
are just the matrix elements of quark currents
between hadron states, which can be extracted assuming isospin symmetry from the leading (non-rare) semileptonic decay $K^+\to\pi^0e^+\nu$ that is
very well measured. Isospin breaking and electroweak corrections are also
 known \cite{Mescia:2007kn}.

The case of $K_{L,S}\to\mu^+\mu^-$ and  $K_{L,S}\to\pi^0\ell^+\ell^-$ is different
as they are subject to long distance contributions. {However, over the past} years
the {understanding} of the latter contributions has been improved by much
\cite{Buchalla:2003sj,Isidori:2003ts,DAmbrosio:2017klp,Mescia:2006jd,DAmbrosio:2018ytt,DAmbrosio:2019xph,Dery:2021mct}. In particular
\begin{itemize}
\item
  It has been demonstrated in
  \cite{Buchalla:2003sj,Isidori:2003ts,Mescia:2006jd} that with the help of the measurements of $K_{S}\to\pi^0\ell^+\ell^-$
  and $K_L\to\pi^0\gamma\gamma$ long distance (LD) contributions to both  $K_{L}\to\pi^0e^+e^-$ and
  $K_{L}\to\pi^0\mu^+\mu^-$ can be determined. Consequently, the short distance (SD)
  contributions to these decays can be extracted from data as well. This is important because
as stressed in \cite{Mescia:2006jd}
$K_L\to\pi^0\mu^+\mu^-$ and $K_L\to\pi^0e^+e^-$ considered simultaneously
offer a powerful test of not only vector and axial-vector currents but in particular of scalar and pseudoscalar currents.
\item
  It has been pointed out in \cite{DAmbrosio:2017klp} that
the SD
parameters of the decay $K\to\mu^+\mu^-$ can be cleanly extracted from a measurement of   the $K_L-K_S$ interference term in the time dependent rate
and consequently to measure   direct CP-violation in this decay.
\item
  Subsequently it has been demonstrated in \cite{Dery:2021mct} that the
 SD  contribution to $K_{S}\to\mu^+\mu^-$ can be extracted from data, making it
  another precision observable.
\item
  But also in the case of $K_L\to\mu^+\mu^-$ significant progress has been
  made so that  already for many years this decay served to bound the estimates for the $\kpn$ rate in various NP scenarios \cite{Isidori:2003ts,DAmbrosio:1996kjn,Gerard:2005yk,GomezDumm:1998gw},
  depending on whether NP contributions are dominated by left-handed or right-handed currents. Explicit models will be listed in the context of our paper.
\end{itemize}

The investigation of these low-energy rare decay processes in
conjunction with their theoretical cleanliness allows to probe,
albeit indirectly, high energy scales of the theory far beyond the reach
of the LHC. They are also very sensitive to the values of the CKM parameters,  in particular
to $V_{td}$ and $\IM\lambda_t= \IM V^*_{ts} V_{td}$ so that the latter could
in principle be extracted
from precise measurements of the decay rates for
$K^+\to\pi^+\nu\bar\nu$ and $K_{\rm L}\to\pi^0\nu\bar\nu$,
respectively.
Moreover, the combination of these two decays offers one of the
cleanest measurements of $\sin 2\beta$  \cite{Buchalla:1994tr} with $\beta$ being one of the angles of the Unitarity Triangle.
However, the very fact
that these processes are based on higher order electroweak effects
implies that their branching ratios are expected to be very small and not easy to access experimentally.

The large sensitivity of the decays in question to the values of the CKM parameters,
in particular $\vcb$, is presently problematic in view of the tensions between
the inclusive and exclusive determinations of this important CKM parameter
\cite{Bordone:2019guc,Bordone:2021oof,Ricciardi:2021shl,Aoki:2019cca}.
Fortunately, as demonstrated recently in \cite{Buras:2021nns}, constructing particular
ratios of the branching ratios for $\kpn$ and $\klpn$ with the parameter
$\varepsilon_K$ allows to remove within the SM the dependence on $\vcb$ entirely
and practically also the one due to $\gamma$ leaving the dependence only on the angles $\beta$ of the unitarity
triangle. As the angle $\beta$ is already well measured through the $S_{\psi K_S}$ asymmetry, this strategy
allows to obtain the SM predictions for both decays that are most accurate to date. It can be applied to other decays
\cite{Buras:2021nns} and we will summarize it in the context of our presentation. Yet, eventually, it will be very important to determine CKM parameters
with the help of tree-level decays because taking ratios could in principle
cancel out NP contributions. Moreover finding an anomaly in a ratio does not yet
tell us in which of the two observables, taking part in the ratio, NP is
present. In fact it could be present in both. We will also discuss this issue below.

As of 2022 one can look back at four decades of theoretical efforts
to calculate the branching ratios for all these decays within the SM.
Among early calculations are \cite{Gaillard:1974hs,Inami:1980fz} in which
QCD corrections were {neglected}. The first LO QCD corrections have been
calculated in \cite{Ellis:1982ve,Dib:1989cc} and the NLO ones in the 1990s
\cite{Buchalla:1993bv,Buchalla:1993wq,Buras:1994qa,Misiak:1999yg,Buchalla:1998ba}. Already the NLO calculations reduced significantly various renormalization scale uncertainties present at LO. Yet, in the last twenty  years further progress has been
made through the following calculations:
\begin{itemize}
\item
NNLO QCD corrections to the
charm contributions in $\kpn$: \cite{Buras:2005gr,Buras:2006gb,Gorbahn:2004my}.
\item
Isospin breaking effects and
non-perturbative effects: \cite{Isidori:2005xm,Mescia:2007kn}.
\item
Complete NLO electroweak corrections to the charm quark contribution to $\kpn$: \cite{Brod:2008ss}.
\item
Complete NLO electroweak corrections to the top quark contribution
to $\kpn$ and $\klpn$: \cite{Brod:2010hi}.
\end{itemize}

As far as $K_{L,S}\to\mu^+\mu^-$ and  $K_L\to\pi^0\ell^+\ell^-$ are concerned
\begin{itemize}
\item
  The NLO QCD calculations have been performed in \cite{Buchalla:1993wq,Buchalla:1998ba} and \cite{Buras:1994qa}, respectively and the NNLO ones for
   $K_{L}\to\mu^+\mu^-$ in \cite{Gorbahn:2006bm}.
\item
  The {long-distance} contributions have been investigated in \cite{Isidori:2003ts,DAmbrosio:2017klp,Mescia:2006jd,DAmbrosio:2018ytt,DAmbrosio:2019xph}.
\end{itemize}
This list of theoretical papers demonstrates very clearly the importance of these decays. Reviews on NLO and NNLO QCD corrections can be found in \cite{Buchalla:1995vs,Buras:2011we}.

The main theoretical uncertainties in the SM predictions for the decays
in question
stem  then from the CKM parameters, in particular $\vcb$ in the case of CP-conserving decays like $\kpn$ and both $\vcb$ and $\vub$ in the case of CP-violating ones like $\klpn$. But further improvements in theory would also be desirable.
In particular in the
case of $\kpn$ an improved estimate of long distance effects in
charm contributions would be welcome. Lattice QCD should be helpful in this respect
and in fact first steps in this direction have been made by {the} RBC-UKQCD
collaboration \cite{Christ:2016eae,Christ:2019dxu}.
It is expected that in the second half of the 2020s the theoretical errors in present estimates of various branching ratios will be significantly reduced.

While, as far as the theory is concerned, the situation of the decays in question is satisfactory within the SM and actually also within a number of BSM
scenarios, this is not the case on the experimental side. The main reason are
the very low branching ratios which require years, even decades to be measured.
In fact the first NLO QCD calculations \cite{Buchalla:1993bv,Buchalla:1993wq,Buras:1994qa}
have been performed almost three decades ago and although the measurements
of the branching ratio for $\kpn$ with the error of $10\%$ is expected
in the coming years, it could still take the full decade to measure the
remaining branching ratios with respectable precision.

It appears then that in this decade the main breakthrough will be made
by experimentalists through the measurements of the six branching ratios
in question and in particular through the measurements of the
distributions in $q^2$, the invariant mass-squared of the neutrino system
in the case of $\kpn$ and $\klpn$ and of the $\ell^+\ell^-$ system
in the case of the remaining decays. These have already been discussed
in particular in \cite{Buchalla:2003sj,Isidori:2003ts,DAmbrosio:2017klp,Mescia:2006jd,DAmbrosio:2018ytt,DAmbrosio:2019xph,Dery:2021mct} and  \cite{Li:2019fhz,Deppisch:2020oyx}   but an improved analysis of them within the
SM and various BSM scenarios would be very desirable so that also theorists
will be able to make further advances in this field in the coming years.
As we will see here also the insight from the Standard Model Effective Field
Theory (SMEFT) will turn out to be useful.

Reviews of these decays can already be found in
\cite{Buras:2020xsm,Buras:2004uu,Komatsubara:2012pn,Buras:2013ooa,Blanke:2013goa,Smith:2014mla} and the power of them in testing
energy scales as high as several hundreds of $\tev$ has been demonstrated
in \cite{Buras:2014zga}. Yet, the presentation here includes also several recent
insights which cannot be found in these papers.

Our paper is organized as follows. In Section~\ref{sec:2} we list the relevant formulae for the branching ratios in question within the SM and
BSM scenarios with left-handed and right-handed quark currents. We
  give also
their estimates within the SM, stressing the issue of parametric CKM uncertainties.
Furthermore, we summarize the experimental status of these decays.
In Section~\ref{sec:2a} the results of \cite{Buras:2021nns} are reviewed,
where the strong $\vcb$ dependence of the branching ratios considered by us
has been eliminated in favour of $\varepsilon_K$.
In Section~\ref{sec:2b} we discuss the differential $q^2$ distributions for all
decays. As stressed two years ago in \cite{Li:2019fhz,Deppisch:2020oyx} and analyzed in detail very recently in \cite{Buras:2022hws} these distributions in
  the case of $\kpn$ and $\klpn$   allow to distinguish between
  Dirac and Majorana neutrinos which is not possible on the basis of
  the branching ratios alone.
In Section~\ref{sec:3} we generalize the discussion
of the previous sections  beyond the SM,
including right-handed currents, scalar/pseudoscalar and tensor operators. This is followed in Section~\ref{sec:4} by the results
obtained in the BSM scenarios stressing the correlations between these decays,
$\epe$,  $\varepsilon_K$ and the $K^0-\bar K^0$ mass difference $\Delta M_K$.
 An outlook is given in Section~\ref{sec:5}.

In the present paper when discussing BSM physics we concentrate
  on heavy particles with masses significantly larger than the electroweak scale. A systematic analysis for the case of light BSM physics, i.e. for NP models with new degrees of freedom lighter than the Kaon mass, can be found in \cite{Goudzovski:2022vbt}.

\section{Basic Formulae}\label{sec:2}
The general formulae for the branching ratios in the presence of left-handed and right-handed vector quark currents within and beyond the SM listed below are sufficient to get an
idea of the structure of various effects and the numerics. Further details,
 in particular the derivations of these formulae,
 can be found in  \cite{Buras:2020xsm} and in the original papers listed there
 and below.

\boldmath
\subsection{ $\kpn$ and $\klpn$} \label{sec:HeffRareKB}
\unboldmath
\boldmath
\subsubsection{ $\kpn$}
\unboldmath
Including isospin breaking corrections,
summing over three neutrino flavours and generalizing the SM formulae in \cite{Buchalla:1998ba, Mescia:2007kn} to include both left-handed and right-handed quark currents one finds
\begin{equation}\label{bkpnn}
\mathcal{B}(K^+\to\pi^+\nu\bar\nu)=\kappa_+ (1+\Delta_\text{EM})\cdot
\left[\left(\frac{{\rm Im}X_{\rm eff}}{\lambda^5}\right)^2+
\left(\frac{{\rm Re}\lambda_c}{\lambda}P_c(X)+
\frac{{\rm Re}X_{\rm eff}}{\lambda^5}\right)^2\right]\,,
\end{equation}
\noindent
where
\be\label{Xeff}
X_{\rm eff}=V_{ts}^*V_{td}\,(X_L(K)+X_R(K))\equiv
V_{ts}^* V_{td} X_L^{\rm SM}(K) ( 1 +\xi e^{i\theta}).
\ee
Here
\be
X_L(K)=X(x_t)+\Delta X_L(K)
\ee
represents contributions of left-handed currents where $\Delta X_L(K)$
denotes BSM contributions and $X(x_t)$ is the SM contribution. $X_R(K)$
represents the contributions of right-handed currents. Only {\em vector} parts  of these currents contribute to these decays. The SM contribution is
given by
\be\label{PCNNLO}
{X(x_t)=1.462\pm0.017, \qquad P_c(X)=0.405\pm 0.024}\,,
\ee
with the latter calculated for $\lambda=0.225$. The value of $X(x_t)$ is the most recent one from  \cite{Brod:2021hsj} corresponding to the most recent value for $m_t(m_t)$ in Table~\ref{tab:input}.
Examples of the BSM contributions for $Z^\prime$ scenarios can be found
in \cite{Buras:2012jb}.

Next
\begin{equation}\label{kapp}
\lambda=\vus,\qquad \kappa_+={ (5.173\pm 0.025 )\cdot 10^{-11}\left[\frac{\lambda}{0.225}\right]^8} , \qquad \Delta_\text{EM}=-0.003.
\end{equation}
Here $x_t=m^2_t/M^2_W$, $\lambda_i=V^*_{is}V_{id}$ are the CKM factors
 and $\kappa_+$ summarizes all the remaining factors, in particular the relevant hadronic matrix element  that can
be extracted from leading semi-leptonic decays of $K^+$, $K_L$ and $K_S$ mesons
\cite{Mescia:2007kn}.
In obtaining the numerical value in~(\ref{kapp})
\be\label{INPUT}
\sin^2\vartheta_W\equiv s_W^2=0.23116, \qquad \alpha(M_Z)=\frac{1}{127.9},
\ee
given  in the $\overline{\text{MS}}$ scheme, have been used.
Their errors are below $0.1\%$ and can be neglected. Further details can be found in Section 9.5.3 of \cite{Buras:2020xsm}.
The SM prediction for the $\kpn$ decay as usually quoted in the literature reads as follows \cite{Buras:2015qea,Bobeth:2016llm}
\be\label{KSM}
\mathcal{B}(\kpn)_\text{SM}= (8.5^{+1.0}_{-1.2})\times 10^{-11}\,, \qquad (2016).
\ee
However, as stressed in \cite{Buras:2021nns}, this result corresponds to
the values of $\vcb$ in the ballpark of inclusive determinations of this parameter and would be significantly lower if the value from exclusive determinations was used. We will return to this important issue in Section~\ref{sec:2a}.

On the experimental side
the NA62 experiment at CERN is  presently running and is expected to measure the $\kpn$ branching ratio with the precision of $10\%$  by 2024, as described in
\cite{Ceccucci:2018bnw,NA62:2020upd}, that would improve the accuracy of the
most recent  measurement by a factor of five.

This measurement from NA62 \cite{CortinaGil:2020vlo, NA62ichep}  reads
\be\label{EXP19}
\mathcal{B}(\kpn)_\text{exp}=(11.0^{+4.0}_{-3.5}\pm 0.3)\times 10^{-11}\,,
\ee
fully consistent with the SM estimates but still leaving room for significant NP contributions.

\begin{table}[H]
\center{\begin{tabular}{|l|l|}
\hline
$m_{B_s} = 5366.8(2)\mev$\hfill\cite{Zyla:2020zbs}	&  $m_{B_d}=5279.58(17)\mev$\hfill\cite{Zyla:2020zbs}\\
$\Delta M_s = 17.749(20) \,\text{ps}^{-1}$\hfill \cite{Zyla:2020zbs}	&  $\Delta M_d = 0.5065(19) \,\text{ps}^{-1}$\hfill \cite{Zyla:2020zbs}\\
{$\Delta M_K = 0.005292(9) \,\text{ps}^{-1}$}\hfill \cite{Zyla:2020zbs}	&  {$m_{K^0}=497.61(1)\mev$}\hfill \cite{Zyla:2020zbs}\\
$S_{\psi K_S}= 0.699(17)$\hfill\cite{Zyla:2020zbs}
		&  {$F_K=155.7(3)\mev$\hfill  \cite{Aoki:2019cca}}\\
	$|V_{us}|=0.2253(8)$\hfill\cite{Zyla:2020zbs} &
 $|\eps_K|= 2.228(11)\cdot 10^{-3}$\hfill\cite{Zyla:2020zbs}\\
$F_{B_s}$ = $230.3(1.3)\mev$ \hfill \cite{Aoki:2021kgd} & $F_{B_d}$ = $190.0(1.3)\mev$ \hfill \cite{Aoki:2021kgd}  \\
$F_{B_s} \sqrt{\hat B_s}=256.1(5.7) \mev$\hfill  \cite{Dowdall:2019bea}&
$F_{B_d} \sqrt{\hat B_d}=210.6(5.5) \mev$\hfill  \cite{Dowdall:2019bea}
\\
 $\hat B_s=1.232(53)$\hfill\cite{Dowdall:2019bea}        &
 $\hat B_d=1.222(61)$ \hfill\cite{Dowdall:2019bea}
\\
{$m_t(m_t)=162.83(67)\gev$\hfill\cite{Brod:2021hsj} }  & {$m_c(m_c)=1.279(13)\gev$} \\
{$S_{tt}(x_t)=2.303$} & {$S_{ut}(x_c,x_t)=-1.983\times 10^{-3}$} \\
    $\eta_{tt}=0.55(2)$\hfill\cite{Brod:2019rzc} & $\eta_{ut}= 0.402(5)$\hfill\cite{Brod:2019rzc}\\
$\kappa_\varepsilon = 0.94(2)$\hfill \cite{Buras:2010pza}	&
$\eta_B=0.55(1)$\hfill\cite{Buras:1990fn,Urban:1997gw}\\
$\tau_{B_s}= 1.515(4)\,\text{ps}$\hfill\cite{Zyla:2020zbs} & $\tau_{B_d}= 1.519(4)\,\text{ps}$\hfill\cite{Zyla:2020zbs}
\\
\hline
\end{tabular}  }
\caption {\textit{Values of the experimental and theoretical
    quantities used as input parameters. For future
updates see FLAG  \cite{Aoki:2021kgd}, PDG \cite{Zyla:2020zbs}  and HFLAV  \cite{Aoki:2019cca}.
}}
\label{tab:input}
\end{table}

\boldmath
\subsubsection{$\klpn$}
\unboldmath
Including isospin breaking corrections in relating
$\klpn$ to $K^+\to\pi^0e^+\nu$ and
summing over three neutrino flavours and generalizing the SM formulae in
\cite{Buchalla:1996fp,Buchalla:1995vs}
to include contributions from both left-handed and right-handed quark currents one finds
\begin{equation}\label{bklpn}
\mathcal{B}(\klpn)=\kappa_L\cdot
\left(\frac{{\rm Im}X_{\rm eff}}{\lambda^5}\right)^2,
\end{equation}
\noindent
where \cite{Mescia:2007kn}
\begin{equation}\label{kapl}
\kappa_L=
(2.231\pm 0.013)\cdot 10^{-10}\left[\frac{\lambda}{0.225}\right]^8\,.
\end{equation}
Due to the absence of $P_c(X)$ in \eqref{bklpn},
$\mathcal{B}(K_L\to\pi^0\nu\bar\nu)$
has essentially no theoretical uncertainties. It is only affected by
parametric uncertainties coming from $m_t$, $\imlt$ and $\kappa_L$ of which
only the one due to $\imlt$ is important.

The SM prediction for $\klpn$ decay usually quoted in the literature is given as follows \cite{Buras:2015qea,Bobeth:2016llm}
\be\label{KSM2}
\mathcal{B}(\klpn)_\text{SM}=(3.2^{+1.1}_{-0.7})\times 10^{-11}\,,\qquad (2016)
\ee
accompanied by the same remarks on $\vcb$ as made after \eqref{KSM}.
The most recent
$90\%$ confidence level (CL) upper bound on $\klpn$  from KOTO  \cite{Ahn:2018mvc} reads
\be\label{EXP19L}
\mathcal{B}(\klpn)_\text{exp}\le 3.0\times 10^{-9}\,.
\ee

The expected
measurement of $\klpn$  by KOTO at J-PARC \cite{Komatsubara:2012pn,Shiomi:2014sfa} should reach the SM level by 2024. Moreover, the KLEVER experiment at CERN SPS \cite{Ambrosino:2019qvz,NA62:2020upd} is expected to measure this decay in this decade.

The KOTO collaboration presented also data on four candidate events in the signal region, finding
\be\label{KOTO}
\mathcal{B}(\klpn)_\text{KOTO}=2.1^{+2.0(+4.1)}_{-1.1(-1.7)}\times 10^{-9}\,,
\ee
at the 68 (95) \% CL. The central  value is by a factor of 65
above the central SM prediction and in fact violates the {Grossman-Nir} bound
\cite{Grossman:1997sk}
  which at the $90\%$ CL together with the present NA62 result for $\kpn$ amounts
  to $0.8\times 10^{-9}$. Theoretical analyses of this interesting data
  can be found in \cite{Kitahara:2019lws,He:2020jly,He:2020jzn,Fuyuto:2014cya}.

\boldmath
\subsubsection{ Interplay of $\kpn$ and $\klpn$}
\unboldmath
Beyond the SM the only unknown  in (\ref{bkpnn}) and (\ref{bklpn})  is the complex function $X_{\rm eff}$. Once both
branching ratios will be measured one day, $X_{\rm eff}$
will be determined model independently as follows  \cite{Buras:2015yca}
\begin{align}
{\rm Re}\,X_{\rm eff} &= -\lambda^5\left[\frac{\B(\kpn)}{\kappa_+(1 + \Delta_{\rm EM})} - \frac{\B(\klpn)}{\kappa_L}\right]^{1/2} - \lambda^4{\rm Re}\,\lambda_c P_c(X)\,,\label{ReX}\\
{\rm Im}\,X_{\rm eff} &= \lambda^5\left[\frac{\B(\klpn)}{\kappa_L}\right]^{1/2}\,.\label{ImX}
\end{align}
In choosing the signs in these formulae it has been  assumed that NP contributions
 do not reverse the sign of the SM functions. For more general expressions
admitting such a possibility see  \cite{Buras:2001af}. At the Grossman-Nir
bound \cite{Grossman:1997sk} the square root in (\ref{ReX}) vanishes.

\begin{figure}[htb]
\centering%
\includegraphics[width=0.7\textwidth]{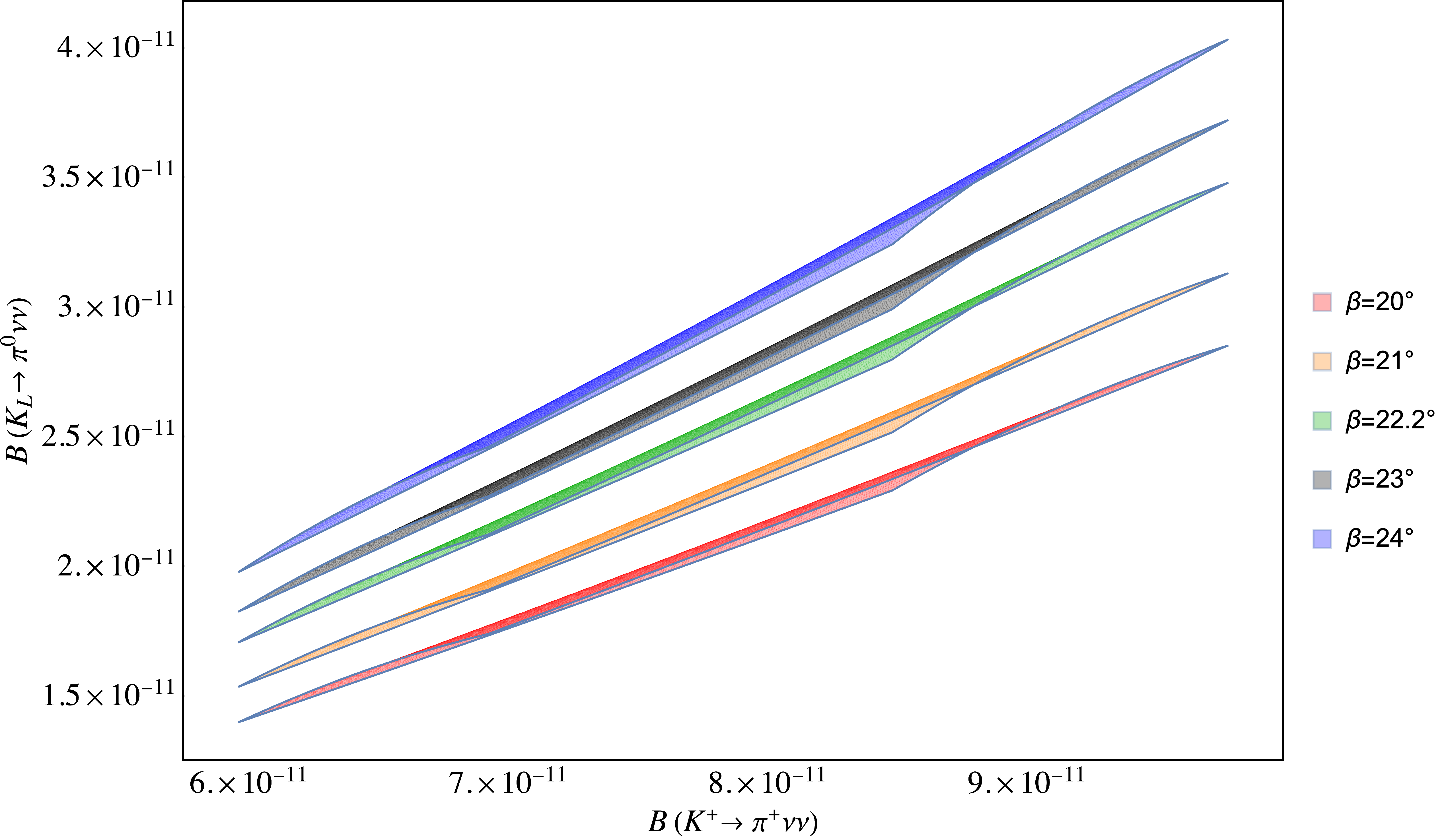}%
\caption{\it The correlation between branching ratios for $\klpn$ and $\kpn$
  for different  values of $\beta$ within the SM.
  The ranges of branching ratios correspond to $38 \leq |V_{cb}|\times 10^{3} \leq 43$ {and $60^\circ\leq\gamma \leq 75^\circ$}. From  \cite{Buras:2021nns}. \label{fig:3}}
\end{figure}

In Fig.~\ref{fig:3} we show correlation between  the branching ratios for
$\klpn$ and $\kpn$ in the SM for fixed values of $\beta$  \cite{Buras:2021nns}.
The SM values depend on $\vcb$ but the positions of the straight lines depend  basically only  on $\beta$ and have negligible dependence on $\gamma$, $m_t$ and $\vcb$ so that they have practically a universal slope. This observation made
already in 1994 in \cite{Buchalla:1994tr} allows one day to determine
the angle $\beta$ from these two branching ratios alone and compare it with the
value extracted from the mixing induced asymmetry $S_{\psi K_S}$.

In this respect an important comment is in order. A given line in  Fig.~\ref{fig:3} reminds us at first sight of the correlation between $\kpn$ and $\klpn$ branching ratios in models with MFV \cite{Buras:2001af}
for $X(x_t)>0$ and in the plots showing this correlations in different
models, like in \cite{Buras:2015yca,Blanke:2009pq}, the SM value is represented by a point.
But one should realize that in those papers the lines are obtained by
varying $X$ while keeping $\vcb$ and $\beta$ fixed. On the other hand in Fig.~\ref{fig:3} while
 $X$ is kept at its SM value both $\vcb$ and $\beta$ are varied. In other
words the SM point in the plots in \cite{Buras:2015yca,Blanke:2009pq} and similar plots   found in the literature is rather uncertain and  Fig.~\ref{fig:3}
signals this uncertainty. Inspecting formulae (\ref{bkpnn}) and (\ref{bklpn}) one finds that for fixed $\beta$ the position on a given straight line in Fig.~\ref{fig:3} is   determined by the combination $\vcb^2 X(x_t)$.
The solution to these large uncertainties has been found recently in
\cite{Buras:2021nns} and we will report on it in  Section~\ref{sec:2a}.

\begin{figure}[htb]
\centering%
\includegraphics[width=0.7\textwidth]{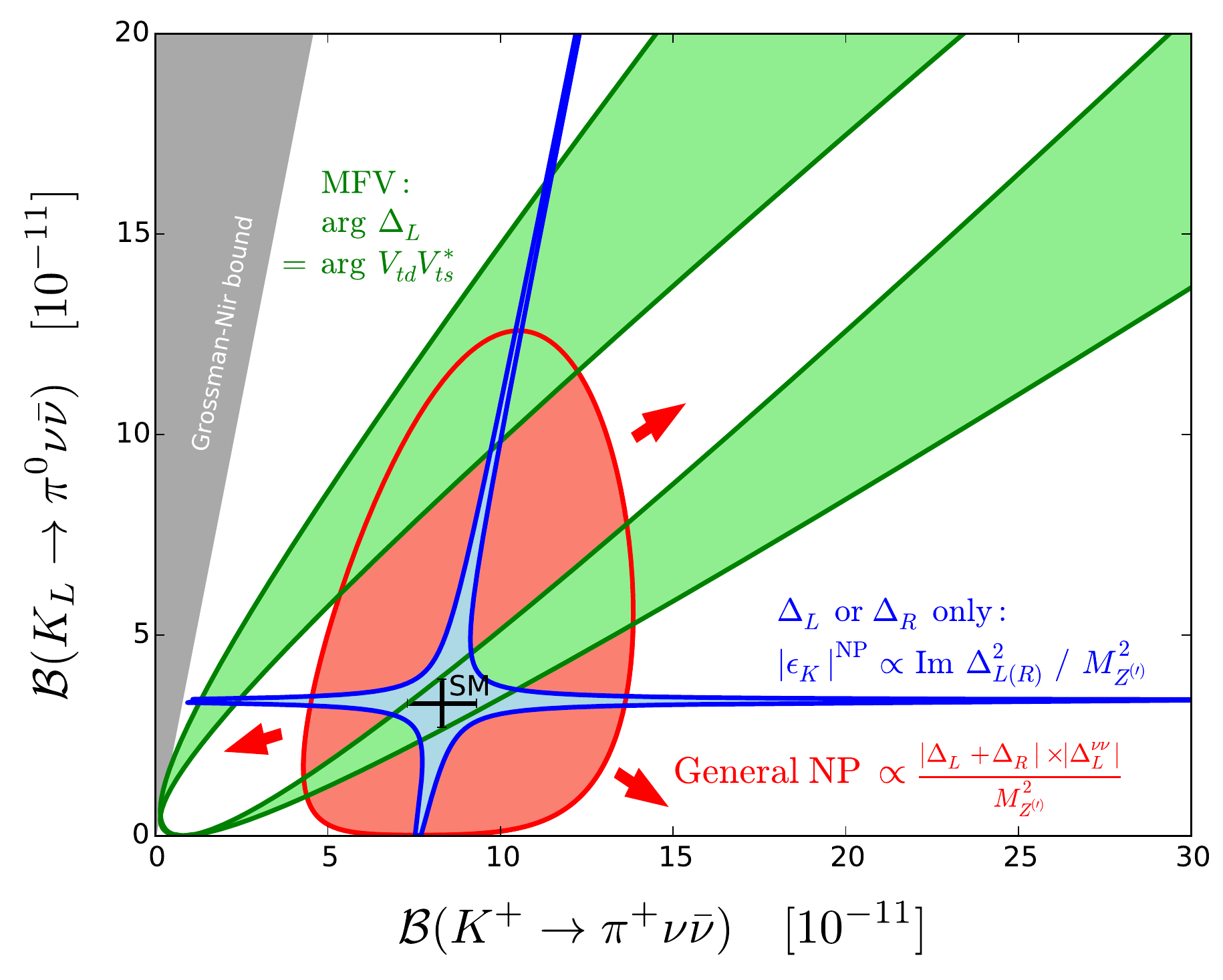}%
\caption{\it Illustrations of common correlations in the $\mathcal{B}(\kpn)$ versus $\mathcal{B}(\klpn)$ plane. The expanding red region illustrates the lack of correlation for models with general LH and RH NP couplings. The green region shows the correlation present in models obeying CMFV. The blue region shows the correlation induced by the constraint from $\varepsilon_K$ if only LH or RH couplings  are present. From \cite{Buras:2015yca}. \label{fig:illustrateEpsK}}
\end{figure}

Postponing the issue of strong $\vcb$ dependence to the next section, let us
summarize what is known about
the correlation between $\mathcal{B}(\kpn)$ and $\mathcal{B}(\klpn)$ in BSM
models. In view of very small hadronic uncertainties this correlation
depends fully on the short distance dynamics, represented by the two real parameters $\xi$ and $\theta$
 {\eqref{Xeff}} that vanish in the SM. Measuring then these branching ratios one day will allow to determine those parameters and, comparing them with their expectations in concrete models, to obtain insight into the flavour structure of the NP contributions. Those can be dominated by left-handed currents, by right-handed currents, or by both with similar magnitudes and phases. In general one can distinguish between three classes of models \cite{Blanke:2009pq}:
\begin{enumerate}
\item
Models with a CKM-like structure of flavour interactions. If based on flavour symmetries only, they include MFV and $U(2)^3$ models \cite{Barbieri:2014tja}.
In this case the function $X_L(K)$ is real and $X_R(K)=0$.
There is then only one variable to our disposal, the value of  $X_L(K)$,  and the only allowed values of both branching ratios are on the  green  branches in figure~\ref{fig:illustrateEpsK}.
But due to stringent correlations with other observables present in this class of models, only certain ranges for $\mathcal{B}(\kpn)$ and $\mathcal{B}(\klpn)$
are still allowed.
\item
Models with new flavour and CP-violating interactions in which either
left-handed currents or right-handed currents fully dominate, implying that
left-right operator contributions to $\varepsilon_K$ can be neglected. In
this case there is a strong correlation between NP contributions to $\varepsilon_K$ and $K\to\pi\nu\bar\nu$ and the $\varepsilon_K$ constraint implies
the blue branch structure shown in figure~\ref{fig:illustrateEpsK}.
On the horizontal branch the NP contribution to $K\to\pi\nu\bar\nu$ is real and therefore vanishes in the case of $\klpn$. On the second branch the NP
contribution is purely imaginary and this branch is parallel to the Grossman-Nir (GN) bound \cite{Grossman:1997sk}. In practice, due to uncertainties in $\varepsilon_K$, there are moderate deviations from this structure which is characteristic for the LHT model \cite{Blanke:2009am}, or $Z$ or $Z^\prime$ FCNC scenarios with either pure LH or RH couplings \cite{Buras:2012jb,Buras:2014zga}.
\item
If left-right operators give a significant contribution to $\varepsilon_K$ or
generally if the correlation between $\varepsilon_K$ and $K\to\pi\nu\bar\nu$
is weak or absent, the two branch structure is also absent. Dependent on
the values of $\xi$ or $\theta$, any value of $\mathcal{B}(\kpn)$ and  $\mathcal{B}(\klpn)$ is in principle possible. The red region in figure~\ref{fig:illustrateEpsK} shows the resulting structure for a fixed value of $\xi$ and $0\le\theta\le 2\pi$. Randall-Sundrum models with
custodial protection belong to this class of models \cite{Blanke:2008yr}.
However, it should be kept in mind that usually the removal of the correlation with $\varepsilon_K$ requires subtle cancellations between different
contributions to $\varepsilon_K$ and consequently some tuning of the parameters \cite{Blanke:2008yr,Buras:2014zga}. This presentation was rather general. We will return to this correlation in explicit models in Section~\ref{sec:4}.
\end{enumerate}

Unfortunately, on the basis of only these two branching ratios alone it is not possible to find out how important the
contributions of right-handed currents are, as their effects are hidden in
a single function $X_{\rm eff}$. In this sense the decays $K_{L,S}\to\mu^+\mu^-$,
$B\to K(K^*)\nu\bar\nu$, as well as $B_{s,d}\to\mu^+\mu^-$
are complementary, and the correlation between $K\to\pi\nu\bar\nu$ decays and
the latter ones can help in identifying the presence or absence of right-handed currents.

Another important issue addressed
  two years ago in \cite{Li:2019fhz,Deppisch:2020oyx} and analyzed in detail very recently in \cite{Buras:2022hws} is the impact of scalar currents on $\kpn$ and
  $\klpn$ decays. In particular such contributions signal the Majorana
  character of neutrinos.
  In the latter paper an anatomy of the
  impact of scalar currents on the correlation between $\kpn$ and $\klpn$ branching ratios has been made. However as emphasized in the latter paper on the
  basis of branching ratios only it is not possible to
  distinguish between    Dirac and Majorana neutrinos. To this end the distributions in $q^2$, the invariant-mass squared of the neutrino pair, are required and we will return to this issue briefly in
  Section~\ref{sec:2b}.

\boldmath
\subsection{$K_{L}\to \mu^+\mu^-$}\label{sec:KLmm}
\unboldmath
Only the so-called SD part of a dispersive contribution
to $K_L\to\mu^+\mu^-$ can be reliably calculated. Despite this limitation,
this contribution puts important bounds on certain NP scenarios.
In contrast to  $\kpn$ and $\klpn$ now instead of the vector
current the axial-vector current contributes.  Relating the relevant
matrix element $\langle 0|\bar s\gamma_\mu P_L d |K_L\rangle$ to the branching
ratio $\mathcal{B}(K^+\to\mu^+\nu_\mu)$ one finds
($\lambda=0.2252$)
\be
{\mathcal{B}(K_L\to\mu^+\mu^-)_{\rm SD} = \kappa_\mu
\left( \frac{{\rm Re}\,Y_{\rm eff}}{\lambda^5} + \frac{{\rm Re}\,\lambda_c}{\lambda} P_c(Y)  \right)^2\,,}
\ee
where
\be
\kappa_\mu = 2.01\cdot 10^{-9}\,, \qquad P_c(Y) = 0.115\pm 0.017
\ee
with $P_c(Y)$ representing the charm contribution at NNLO \cite{Gorbahn:2006bm}.

Similar to \eqref{Xeff}
\be\label{Yeff}
Y_{\rm eff} = V_{ts}^* V_{td} \,(Y_L(K)-Y_R(K)),
\ee
except for the explicit minus sign that has been  introduced to emphasize
that this decay is governed by axial-vector currents as opposed
to $K\to\pi\nu\bar\nu$ decays governed by vector-currents.

Here
\be
Y_L(K)=Y(x_t)+\Delta Y_L(K), \qquad Y(x_t)=0.942\,,
\ee
represent contributions of left-handed currents with $\Delta Y_L(K)$
denoting BSM contributions and $Y(x_t)$ being the SM contribution at the NNLO \cite{Bobeth:2013tba}. $Y_R(K)$
represents the contributions of right-handed currents.
Examples of the BSM contributions for $Z^\prime$ scenarios can be found
in \cite{Buras:2012jb}.

We find then
\be
\mathcal{B}(K_L\to\mu^+\mu^-)^\text{SM}_{\rm SD}\approx  (0.8\pm0.1)\cdot 10^{-9}\,.
\ee
The extraction of the short distance
part from data is subject to considerable uncertainties.
Here the important issue is the  sign of the interference of the SD dispersive
part $\chi_{\rm SD}$ of the decay amplitude of $K_L \to \mu\bar\mu$ with the
corresponding LD parts. Allowing for both signs implies a conservative bound $|\chi_{\rm SD}| \leq 3.1$ \cite{Isidori:2003ts}. This gives then the known
upper bound \cite{Isidori:2003ts}
\be\label{eq:KLmm-bound}
\mathcal{B}(K_L\to\mu^+\mu^-)_{\rm SD} \le 2.5 \cdot 10^{-9}\,,
\ee
roughly three times as large as the SM value.
This bound is also obtained for the sign favoured in
\cite{DAmbrosio:1996kjn,Gerard:2005yk} that implies $-1.7 \leq \chi_{\rm SD}\leq 3.1$.

On the
other hand the opposite sign is favoured in~\cite{GomezDumm:1998gw}, giving  $-3.1 \leq \chi_{\rm SD}\leq 1.7$  and therefore approximately
\be\label{eq:KLmm-bound1}
\mathcal{B}(K_L\to\mu^+\mu^-)_{\rm SD} \le \mathcal{B}(K_L\to\mu^+\mu^-)^{\rm SM}_{\rm SD}\,.
\ee
The implications of these bounds will be discussed in  Section~\ref{sec:4}.
 We will find there that they do not allow large enhancements of $\mathcal{B}(\kpn)$ for models with NP governed by left-handed currents but are much less
 important if right-handed currents dominate NP contributions.

 More recently, it has been pointed out in \cite{DAmbrosio:2017klp} that the
SD
parameters of the decay $K\to\mu^+\mu^-$ can be cleanly extracted from a measurement of the $K_L-K_S$ interference term in the time-dependent rate
and consequently direct CP-violation can be measured in this decay.
This brings us to the next even more interesting decay.

\boldmath
\subsection{$K_{S}\to \mu^+\mu^-$}\label{sec:KSmm}
\unboldmath
The decay $\ksm$ provides a  sensitive probe of imaginary parts of
short-distance couplings. Its branching fraction receives LD and
SD contributions, which are added incoherently in the total
rate \cite{Ecker:1991ru, Isidori:2003ts}. This is in contrast to the decay
$\klm$, where LD and SD amplitudes interfere and moreover $\BR(\klm)$ is
sensitive to the real parts of couplings. The SD part of $\BR(\ksm)$ is given as
\begin{equation}
  \label{eq:ksm-br-SD}
 \BR(\ksm)_{\rm SD}
  = \tau_{K_S} \frac{G_F^2 \alpha^2}{8 \pi^3\sin^4\theta_W} m_K F_K^2 \sqrt{1-\frac{m^2_\mu}{m_K^2}} m_\mu^2\,
    \left[{\rm Im} Y_{\rm eff} \right]^2=1.04\times 10^{-5}\left[{\rm Im} Y_{\rm eff} \right]^2,
\end{equation}
with $ Y_{\rm eff}$ given in (\ref{Yeff}).

In 2017 the LHCb collaboration improved the upper bound on $\ksm$
by one order of magnitude \cite{Aaij:2017tia}
\begin{align}
  \label{ksmbound}
  \BR(\ksm)_{\rm LHCb} & < 0.8\, (1.0) \times 10^{-9} &
  \mbox{at} \; 90\%\, (95\%) \; \mbox{C.L.}
\end{align}
to be compared with the SM prediction \cite{Isidori:2003ts, DAmbrosio:2017klp}
\begin{align}\label{ISDA}
  \BR(\ksm)_{\rm SM} &
  = (4.99_{\rm LD} + 0.19_{\rm SD}) \times 10^{-12}
  = (5.2 \pm 1.5) \times 10^{-12}.
\end{align}
There are good future prospects to improve this bound, LHCb expects
\cite{Cerri:2018ypt} with 23~fb$^{-1}$ sensitivity to regions
$\BR(\ksm) \in [4,\, 200] \times 10^{-12}$, close to the SM prediction.
 As already mentioned previously it has been demonstrated in \cite{Dery:2021mct} that the short distance
  contribution to $K_{S}\to\mu^+\mu^-$ can be extracted from data, making it
  another precision observable.

  This is important because this decay being dominated by direct CP-violation
  in models with axial-vector currents is very sensitive to NP contributions
  as recently analysed in \cite{Dery:2021vql} but also earlier. See
  in particular analyses of $Z^\prime$ models \cite{Buras:2012jb}, leptoquark models \cite{Bobeth:2017ecx} and several models reviewed in \cite{Buras:2020xsm}.

\boldmath
\subsection{$K_L\to\pi^0 \ell^+\ell^-$}\label{sec:KLpmm}
\unboldmath
The rare decays $K_L\to\pi^0e^+e^-$ and $K_L\to\pi^0\mu^+\mu^-$ are
dominated by CP-violating contributions. In the SM the main
contribution comes from the indirect (mixing-induced) CP violation and
its interference with the direct CP-violating contribution
\cite{D'Ambrosio:1998yj,Buchalla:2003sj,Isidori:2004rb,Friot:2004yr}.
The direct
CP-violating  contribution to the branching ratio is within the SM
in the ballpark of
$4\cdot 10^{-12}$, while the CP conserving contribution is at
most $3\cdot 10^{-12}$. Among the rare $K$ meson decays, the decays in
question belong to the theoretically cleanest, but certainly cannot
compete with the $K\to\pi\nu\bar\nu$ decays. Moreover,
the dominant indirect CP-violating contributions are practically
determined by the measured decays $K_S\to\pi^0\ell^+\ell^-$ and the
parameter $\varepsilon_K$. Consequently they are not as sensitive as the
$K_L\to\pi^0\nu\bar\nu$ decay to NP
 contributions, present only in the subleading direct
CP violation. However, in
the presence of large new CP-violating phases, the direct CP-violating
contribution can become the dominant contribution and the branching
ratios for $K_L\to\pi^0\ell^+\ell^-$ can be enhanced significantly,
 with a stronger effect in the case of
 $K_L\to\pi^0\mu^+\mu^-$ as already analyzed in
 \cite{Isidori:2004rb,Friot:2004yr,Mescia:2006jd}. But what is even more important are the correlations of these decays with $\klpn$ and the ratio $\epe$ which we will encounter in  Section~\ref{sec:4}.

 The expressions for the branching ratios are now more complicated because
 more operators enter the analysis
 and it is better to use the Wilson coefficients of involved operators
 than generalizing the one loop functions to include NP contributions.

 We follow here \cite{Bobeth:2017ecx} where the formulae in \cite{Buchalla:2003sj, Isidori:2004rb, Friot:2004yr,Mescia:2006jd} have been generalized to include NP contributions.

 The rare decays in question are described by the general $\Delta F = 1$
Hamiltonian of the semi-leptonic FCNC transition of down-type quarks into
leptons  below the electroweak (EW) scale $\muEW$
\begin{align}
  \label{eq:DF1-eff-H}
  {\cal H}_{d\to d(\ell \ell,\nu\nu)} & =
  -\frac{4 G_F}{\sqrt{2}} \lambda^{ji}_t \frac{\alpha_e}{4 \pi}
   \sum_k C_k^{baji} Q_k^{baji} + \mbox{h.c.} \,
\end{align}
with $a,b$ being lepton indices and  $i,j$ down-quark indices.
There are eight semi-leptonic operators relevant for $d_i \ell_a \to d_j \ell_b$
when considering UV completions that give rise to the SMEFT above the electroweak
scale \cite{Alonso:2014csa}
\begin{equation}
\begin{aligned}
  Q_{9(9')}^{baji} &
  = [\bar{d}_j \gamma_\mu P_{L(R)} d_i] [\bar{\ell}_b \gamma^\mu \ell_a] ,
& \qquad
  Q_{10(10')}^{baji} &
  = [\bar{d}_j \gamma_\mu P_{L(R)} d_i] [\bar{\ell}_b \gamma^\mu \gamma_5 \ell_a] ,
\\
  Q_{S(S')}^{baji} &
  = [\bar{d}_j P_{R(L)} d_i] [\bar{\ell}_b \ell_a] ,
& \qquad
  Q_{P(P')}^{baji} &
  = [\bar{d}_j P_{R(L)} d_i] [\bar{\ell}_b \gamma_5 \ell_a] .
\end{aligned}
\end{equation}

The SM contribution to these Wilson coefficients is lepton-flavour diagonal
\begin{align}
  C_k^{baji} &
  = C_{k,{\rm SM}} \, \delta_{ba}
  + \frac{\pi}{\alpha_e} \frac{v^2}{\lambda_t^{ji}} \, C_{k,{\rm NP}}^{baji}\,,
\end{align}
where $v=246\gev$ and a normalisation factor has been introduced for the NP
contribution that proves convenient for matching calculations in the SMEFT.
The non-vanishing SM contributions
\begin{align}
  C_{9,{\rm SM}} &
  = \frac{Y(x_t)}{s_W^2} - 4 Z(x_t) \,, &
  C_{10,{\rm SM}} &
  = - \frac{Y(x_t)}{s_W^2} \,, &
  C_{L,{\rm SM}} &
  = - \frac{X(x_t)}{s_W^2} \,,
\end{align}
are given by the gauge-independent functions $X(x_t)$, $Y(x_t)$, which we encountered in previous decays and
$Z(x_t)$ is an additional one-loop function that has to be included due
to the presence of QED penguins that do not contribute to the previous decays~\cite{Buchalla:1990qz}. Here
$s_W \equiv \sin \theta_W$.

Generalising in particular the formulae in \cite{Mescia:2006jd} to include NP contributions and adapting them to our notations
one finds, dropping scalar and pseudoscalar contributions \cite{Bobeth:2017ecx}
\begin{align}
  \label{eq:BrKpiLL}
  \BR(\klpll) &
  = \left( C_\text{dir}^\ell \pm C_\text{int}^\ell \left|a_s\right|
         + C_\text{mix}^\ell \left|a_s\right|^2
         + C_\text{CPC}^\ell \right)\times 10^{-12} \,.
\end{align}
Here \cite{Mescia:2006jd}
\begin{equation}
  \label{eq:klpll-num-1}
\begin{aligned}
  C_\text{dir}^e   & = (4.62\pm0.24)[(\omega_{7V}^e)^2 + (\omega_{7A}^e)^2]\, ,  & \qquad
  C_\text{int}^e   & = (11.3\pm0.3) \, \omega_{7V}^e \,, &
\\
  C_\text{dir}^\mu & = (1.09\pm0.05)[(\omega_{7V}^\mu)^2 + 2.32 (\omega_{7A}^\mu)^2] \,, &
  C_\text{int}^\mu & = (2.63\pm0.06) \, \omega_{7V}^\mu \,,
\end{aligned}
\end{equation}
and
\begin{equation}
  \label{eq:klpll-num-2}
\begin{aligned}
  C_\text{mix}^e   & = 14.5\pm0.05 \,, & \qquad
  C_\text{CPC}^e   & \simeq 0 \,, &      \qquad
  \left|a_s\right| & = 1.2 \pm0.2 \,,
\\
  C_\text{mix}^\mu & = 3.36\pm0.20 \,, &
  C_\text{CPC}^\mu & = 5.2\pm1.6 \,. &
\end{aligned}
\end{equation}
The SM and NP contributions enter through
\begin{align}
  \omega_{7V}^\ell &
  = \frac{1}{2\pi}\left(P_0+ C_{9,{\rm SM}} \right)
    \left[ \frac{{\rm Im}\lambda_t^{sd}}{1.407 \times 10^{-4}} \right]
  + \frac{1}{\alpha_e} \frac{v^2}{2} \frac{{\rm Im} \left[
          C_{9,{\rm NP}}^{\ell\ell sd} + C_{9',{\rm NP}}^{\ell\ell sd}
    \right]}{1.407 \times 10^{-4}}\,,\label{omega7V}
\\
  \omega_{7A}^\ell &
  = \frac{1}{2\pi} C_{10,{\rm SM}}
    \left[ \frac{{\rm Im}\lambda_t^{sd}}{1.407 \times 10^{-4}} \right]
  + \frac{1}{\alpha_e} \frac{v^2}{2} \frac{{\rm Im} \left[
          C_{10,{\rm NP}}^{\ell\ell sd} + C_{10',{\rm NP}}^{\ell\ell sd}
    \right]}{1.407 \times 10^{-4}}\,,\label{omega7A}
\end{align}
where $P_0=2.88\pm 0.06$ \cite{Buras:1994qa} includes NLO QCD corrections and $\ell$ either $e$ or $\mu$.

 NP contributions do not depend on
$\lambda^{sd}_t$ but the factor $1.407 \times 10^{-4}$ is present because it has
been used in \cite{Mescia:2006jd} to obtain the numbers in
\refeq{eq:klpll-num-1} and \refeq{eq:klpll-num-2}.

The effect of NP contributions with vector and axial-vector currents  is mainly felt in
$\omega_{7A}$, as the corresponding contributions in $\omega_{7V}$
cancel each other to a large extent. The case of scalar and pseudoscalar contributions is different and is briefly mentioned below.

The present experimental bounds
\be
\mathcal{B}(K_L\to\pi^0e^+e^-)<28\cdot10^{-11}\quad\text{\cite{AlaviHarati:2003mr}}\,,\qquad
\mathcal{B}(K_L\to\pi^0\mu^+\mu^-)<38\cdot10^{-11}\quad\text{\cite{AlaviHarati:2000hs}}
\ee
are still by one order of magnitude larger than the SM predictions
\cite{Mescia:2006jd}
\begin{gather}
\mathcal{B}(K_L\to\pi^0e^+e^-)_\text{SM}=
3.54^{+0.98}_{-0.85}\left(1.56^{+0.62}_{-0.49}\right)\cdot 10^{-11}\,,\label{eq:KLpee}\\
\mathcal{B}(K_L\to\pi^0\mu^+\mu^-)_\text{SM}= 1.41^{+0.28}_{-0.26}\left(0.95^{+0.22}_{-0.21}\right)\cdot 10^{-11},\label{eq:KLpmm}
\end{gather}
 with the values in parentheses corresponding to the ``$-$'' sign
in \eqref{eq:BrKpiLL}, that is
the destructive interference
between direct and indirect CP-violating contributions.
The last discussion  of the theoretical status of this interference
sign can be found in \cite{Prades:2007ud}, where the results of \cite{Isidori:2004rb,Friot:2004yr,Bruno:1992za} are
critically analysed. From this discussion, constructive interference
seems to be  favoured though more work is necessary. In view of significant
uncertainties in the SM prediction it is common to use these decays
to test whether the correlations of them with $\klpn$ and $\kpn$ decays in
various NP scenarios
can have an impact on the latter decays. In any case there is still much
room for NP contributions in these decays. For a recent theoretical
study of long distance aspects of these decays including
$K^\pm\to\pi^\pm\ell^+\ell^-$ and $K_S\to\pi^0\ell^+\ell^-$
see \cite{DAmbrosio:2018ytt,DAmbrosio:2019xph}.

Detailed numerical analyses of these formulae have been presented in
  \cite{Mescia:2006jd} and in a number of papers listed in Section~\ref{sec:3}.
  In particular the correlation between branching ratios for the $\pi^0e^+e^-$ and $\pi^0\mu^+\mu^-$ channels can be found in Figs. 2 and 4 of that paper.

  As far as scalar and pseudoscalar operators are concerned they do not play
  any role for $\pi^0e^+e^-$ but are larger for  $\pi^0\mu^+\mu^-$. Yet, the
  general conclusion of \cite{Mescia:2006jd} is that the effects of these
  operators are expected to be much smaller than from vector and axial-vector operators.   This finding has been confirmed much later in \cite{Buras:2013rqa}.
As the measurements of both branching ratios will begin to be of interest
  only in the second half of this decade we only show in Fig.~\ref{fig:LHS_kp_k0} and \ref{fig:epsK} the correlation
  of both branching ratios with each other and with branching ratios for
  $\kpn$ and $\klpn$ including only  vector and axial-vector operators.

\boldmath
\section{$\vcb$-Independent Ratios}\label{sec:2a}
\unboldmath
Already many years ago suggestions have been put forward to eliminate at least
approximately the dependence on $\vcb$ from phenomenology of rare decays
in the SM due to strong dependence of branching ratios on $\vcb$, in particular
in $K$ decays but also $B$ decays \cite{Buchalla:1994tr,Buras:2003td}.
With improved experiments and theory these ideas could be applied
in $B$ physics in \cite{Bobeth:2021cxm} and generalized to many rare $B$ and $K$
decays in \cite{Buras:2021nns}. In the latter reference 16 $\vcb$-independent
ratios have been presented and  summarized in Table~4 of that paper.
These relations are not only independent of $\vcb$ but are theoretically very
clean and depend generally only on the angles $\beta$ and $\gamma$ of the unitarity
triangle. In certain cases they depend only on $\beta$ or on $\gamma$ and
sometimes they are CKM independent. Table 4 in \cite{Buras:2021nns} summarizes
these dependences. The angle $\beta$ is known already with good precision and
$\gamma$ should be measured with a precision of $1^\circ$ by LHCb and Belle II in the coming years. As these 16 relations are specific for the SM
a pattern of possible future violations of these relations will
give us very useful hints for the type of NP that influences
rare $K$ and $B$ decays as well as quark mixing.
Here we
will concentrate on those relations that deal entirely with the Kaon system.

To this end it is useful to define  the ``reduced'' branching ratios \cite{Buchalla:1994tr}
\begin{equation}\label{b1b2}
B_1=\frac{\mathcal{B}(\kpn)}{\kappa_+(1+\Delta_{\text{EM}})}\,,\qquad
B_2=\frac{\mathcal{B}(\klpn)}{\kappa_L}\,.
\end{equation}
We find then
\be\label{B1vsB2}
B_1=B_2\left[1+\frac{1}{\sigma^2}(\cot\beta+\frac{\sqrt{\sigma} P_c(X)}{\sqrt{B_2}})^2\right]\,, \qquad \sigma = \left(\frac{1}{1- \frac{\lambda^2}{2}} \right)^2\,.
\ee

This relation summarizes analytically the correlation in Fig.~\ref{fig:3}.
We stress again that to an excellent approximation this relation is independent of $\vcb$, $\gamma$ and $m_t$. Therefore it can  be used in principle to determine the angle $\beta$, the sole parameter in this formula \cite{Buchalla:1994tr}.

{Alternatively one can derive a more transparent formula \cite{Buras:2021nns}.
{\begin{align}
\mathcal{B}(\kpn) &= {(7.92\pm 0.30)}\times 10^{-11}   \left[\frac{\sin 22.2^\circ}{\sin \beta}\right]^{1.4} \left[\frac{\mathcal{B}(\klpn)}{{2.61}\times 10^{-11}}\right]^{0.7}\label{master0}.
\end{align}}
This formula reproduces (\ref{B1vsB2}) with an accuracy in the ballpark of
4\%. Consequently the ratio
\be
{R_0=\frac{\mathcal{B}(\kpn)}{\mathcal{B}(\klpn)^{0.7}}\, ,}
\label{eq:R0}
\ee
is approximately $\vcb$-independent.
Restricting the value of $\beta$ to the PDG value from $S_{\psi K_S}=22.2(7)^\circ$, and including all other uncertainties one finds \cite{Buras:2021nns}
\be
  (R_0)_{\rm SM}={(2.03\pm 0.11)}\times 10^{-3}\,.
  \ee

Of particular interest are also the following relations \cite{Buras:2021nns}
\be\label{R11}
  {R_{11}=\frac{\mathcal{B}(\kpn)}{|\varepsilon_K|^{0.82}}=(1.31\pm0.05)\times 10^{-8}\left(\frac{\sin 22.2^\circ}{\sin \beta}\right)^{0.71}{\left(\frac{\sin\gamma}{\sin 67^\circ}\right)^{0.015}},}
  \ee
  \be\label{R12a}
{R_{12}=\frac{\mathcal{B}(\klpn)}{|\varepsilon_K|^{1.18}}=(3.87\pm0.06)\times 10^{-8}
    \left(\frac{\sin\beta}{\sin 22.2^\circ}\right)^{0.9{8}} {\left(\frac{\sin\gamma}{\sin 67^\circ}\right)^{0.03}}}\,,
  \ee
  and
  \be
  \frac{\mathcal{B}(\klpn)}{\mathcal{B}(\kpn)}={(2.95\pm 0.12)}\,|\varepsilon_K|^{0.36}
  \left(\frac{\sin\beta}{\sin 22.2^\circ}\right)^{1.6{9}} {\left(\frac{\sin\gamma}{\sin 67^\circ}\right)^{0.015}}\,.
  \ee

The first two of these formulae express explicitly the fact that combining
  on the one hand $\kpn$ and $\varepsilon_K$ and on the other hand
  $\klpn$ and  $\varepsilon_K$ allows within the SM to determine
  to a very good approximation the angle $\beta$ independently of the value
  of $\vcb$ and $\gamma$. The last one just follows from them. Indeed
  the dependence on $\gamma$ is very weak. An important test
    will be whether these two determinations of $\beta$ will agree with each other.

    In obtaining these formulae it was important to use the results from
\cite{Brod:2019rzc} where
the significant QCD uncertainty from the pure charm contribution to $\varepsilon_K$ has been practically removed through a clever but simple trick by using CKM unitarity differently than done until now in the literature. Moreover, the inclusion of the two-loop electroweak effects in the top contribution further increases the precision in evaluating $\varepsilon_K$  \cite{Brod:2021qvc}.

  Assuming then no NP in $\varepsilon_K$ allows to find the most accurate SM predictions
  for $\kpn$ and $\klpn$ branching ratios to date  \cite{Buras:2021nns}
    \be\label{KSM3}
\boxed{\mathcal{B}(\kpn)_\text{SM}=(8.60\pm0.42)\times 10^{-11}\,,\quad
\mathcal{B}(\klpn)_\text{SM}=(2.94\pm 0.15)\times 10^{-11}\,.}
\ee
It should be emphasized that these predictions are $\vcb$-independent
and to an excellent accuracy $\gamma$-independent. But they obviously
depend on the assumption that NP contributions to $\varepsilon_K$ are absent.

Comparing with (\ref{KSM})
and (\ref{KSM2}) we note a very strong reduction of the error, in particular
 to obtain (\ref{KSM}) and (\ref{KSM2}) values of $\vcb$ in the ballpark
of inclusive one have been used, which are controversial these days.

Similar one finds  \cite{Buras:2021nns}
\be\label{SR1}
{R_{\rm SL}=\frac{\BR(\ksm)_{\rm SD}}{\BR(\klpn)}=1.55\times 10^{-2}\,\left[\frac{\lambda}{0.225}\right]^2
\left[\frac{Y(x_t)}{X(x_t)}\right]^2\,,}
\ee
to be independent of any SM parameter except for $m_t$ and $\lambda$ which are both precisely known. Consequently using \eqref{KSM3} one finds
\be\label{ksm3}
\BR(\ksm)_{\rm SD}=(1.85\pm 0.10)\times 10^{-13}\,.
\ee

It should be emphasized that in obtaining the results in (\ref{KSM3}) and
(\ref{ksm3}) only the absence of NP contributions to $\varepsilon_K$ and the mixing-induced CP-asymmetry $S_{\psi K_S}$ has been assumed. No assumption about
the absence of NP in $B$ decays and in mass differences $\Delta M_{s,d}$ has been made  and
importantly no global fit to obtain these results was necessary. As stressed by  the authors of \cite{Buras:2021nns} this allows to avoid pollution
from NP and larger theoretical uncertainties in a number of $B$ decays
that are necessarily included in a global fit.

\boldmath
\section{$q^2$ distributions}\label{sec:2b}
\unboldmath
Of particular importance are differential $q^2$ distributions with $q^2$ being
the invariant mass-squared of the neutrino system
in the case of $\kpn$ and $\klpn$ and of the $\ell^+\ell^-$ system
in the case of  the $K_L\to\pi^0\ell^+\ell^-$ decays. They are crucial
for the separation of NP vector current contributions from the scalar ones
which cannot be done on the basis of the branching ratios alone.

\boldmath
\subsection{$\kpn$ and $\klpn$}
\unboldmath
The first studies of $q^2$ distributions
in $\kpn$ and $\klpn$ have been performed in  \cite{Li:2019fhz,Deppisch:2020oyx} and analyzed in great detail very recently in \cite{Buras:2022hws}. Our presentation here, based on the latter paper, is very brief and is intended only
to give an of idea how these distributions look like. The interest
in these distributions originates from the fact that they can distinguish
between {\em vector} current and {\em scalar} current contributions.
As pointed out in  \cite{Li:2019fhz,Deppisch:2020oyx} the presence of scalar
current contributions would be a hint for neutrinos being Majorana particles
while the usual vector contributions represent Dirac neutrinos.

Beginning with the vector current contributions, assuming
 neutrino flavour universality,  summing over neutrino flavour and adjusting the notation in  \cite{Li:2019fhz,Deppisch:2020oyx}
and \cite{Buras:2022hws} to ours we find for vector-currents
\be\label{D1}
{\left[\frac{d\Gamma(K^+\rightarrow \pi^+\nu\bar\nu)}{d q^2}\right]_V=
\frac{r^2}{2^{9}\pi^3m_{K^+}^3} \left | \lambda_c X_{\rm NNL} + X_{\rm eff}(K))\right |^2 \lambda^{3/2}(q^2,m_{K^+}^2,m_{\pi^+}^2)|f^{K^+}_+(q^2)|^2}\,,
\ee
\be\label{D2}
\left[\frac{d\Gamma(K_L\rightarrow \pi^0\nu\bar\nu)}{d q^2}\right]_V=
\frac{r^2}{2^{9}\pi^3m_{K_L}^3}
\left ({\IM}X_{\rm eff}(K) \right )^2\,
\lambda^{3/2}(q^2,m_{K_L}^2,m_{\pi^0}^2)|f^{K_L}_+(q^2)|^2 \,,
\ee
with $X_{\rm eff}(K)$ defined in (\ref{Xeff}),
$\lambda(a,b,c)= a^2  +b^2 + c^2 - 2 (a b+ b c + a c)$ and
\be
r=\frac{G_F}{\sqrt{2}}\frac{2\alpha}{\pi \sin^2\theta_W}\,, \qquad  X_{\rm NNL}=\lambda^4 P_c(X)\,.
\ee
In order to be consistent with (\ref{KSM3}) for the SM contribution
  we use the values in (\ref{PCNNLO}) and
\be
V_{cs}^*V_{cd}= -0.219, \qquad  V_{ts}^*V_{td}=-0.20 \vcb^2 \exp{(-i23^\circ)},\qquad
\vcb=41.8\times 10^{-3}\,.
\ee
The form factor arising from the quark vector current is given by \cite{Mescia:2007kn},
\begin{align}
\label{eq:kaonformfactors1}
	f^K_+(q^2) = f^K_+(0)\left(1 + \lambda_+'\frac{q^2}{m_\pi^2} + \lambda_+''\frac{q^4}{{2}m_\pi^4}\right),
\end{align}
with
\begin{align}
\label{eq:fK0}
	f^{K^+}_+(0) = 0.9778, \quad f^{K_L}_+(0) = 0.9544,
\end{align}
and
$\lambda_+' = 24.82\times10^{-3}$, $\lambda_+'' = 1.64\times10^{-3}$.
In Fig.~\ref{fig:distnunu} we show the distributions (\ref{D1}) and  (\ref{D2}) for the SM. They are
represented by blue dashed lines. We have checked that integrating over $q^2$ reproduces
  the SM branching ratios in (\ref{KSM3}). For scalar current contributions we find using the formalism in
\cite{Li:2019fhz,Deppisch:2020oyx} and  \cite{Buras:2022hws}
\be\label{D3}
\left[\frac{d\Gamma(K^+\rightarrow \pi^+\nu\bar\nu)}{d q^2}\right]_S= |N^+_S|^2
q^2\lambda^{1/2}(q^2,m_{K^+}^2,m_{\pi^+}^2){|f^{K^+}_0(q^2)|^2}\,,
\ee
\be\label{D4}
\left[\frac{d\Gamma(K_L\rightarrow \pi^0\nu\bar\nu)}{d q^2}\right]_S=
|N^L_S|^2 q^2\lambda^{1/2}(q^2,m_{K_L}^2,m_{\pi^0}^2){|f^{K_L}_0(q^2)|^2} \,,
\ee
with scalar formfactors given by \cite{Mescia:2007kn}
\be
\label{eq:kaonformfactors3}
f^{K^+}_0(q^2)= f_+^{K_+}(0)\left(1 + \lambda_0\frac{q^2}{m_{\pi^+}^2} \right), \qquad
f^{K_L}_0(q^2)= f_+^{K_L}(0)\left(1 + \lambda_0\frac{q^2}{m_{\pi^0}^2} \right),
\ee
$f^{K^+}_+(0)$ and $f^{K_L}_+(0)$ given in (\ref{eq:fK0}) and
$\lambda_0=13.38\times 10^{-3}$.
In the left panel of Fig.~\ref{fig:distnunu} we compare  the scalar and vector
contributions for $\kpn$ (left) and for $\klpn$ (right).
The coefficients {$N^+_S$ and $N^L_S$} have been chosen such that the scalar
and vector distributions are of comparable size.
{Note that we have presented our results in the center of mass frame of the Kaon for both $d\Gamma/dq^2$ as well as $\Gamma_{\rm tot}$, in contrast with Fig. 6 of \cite{Deppisch:2020oyx} where the former quantity is in the Lab frame. After accounting for the boost factor, for $\kpn$ our results agree with Fig. 6 of \cite{Deppisch:2020oyx}.}

\begin{figure}[htb]
\centering
\includegraphics[width=0.49\textwidth]{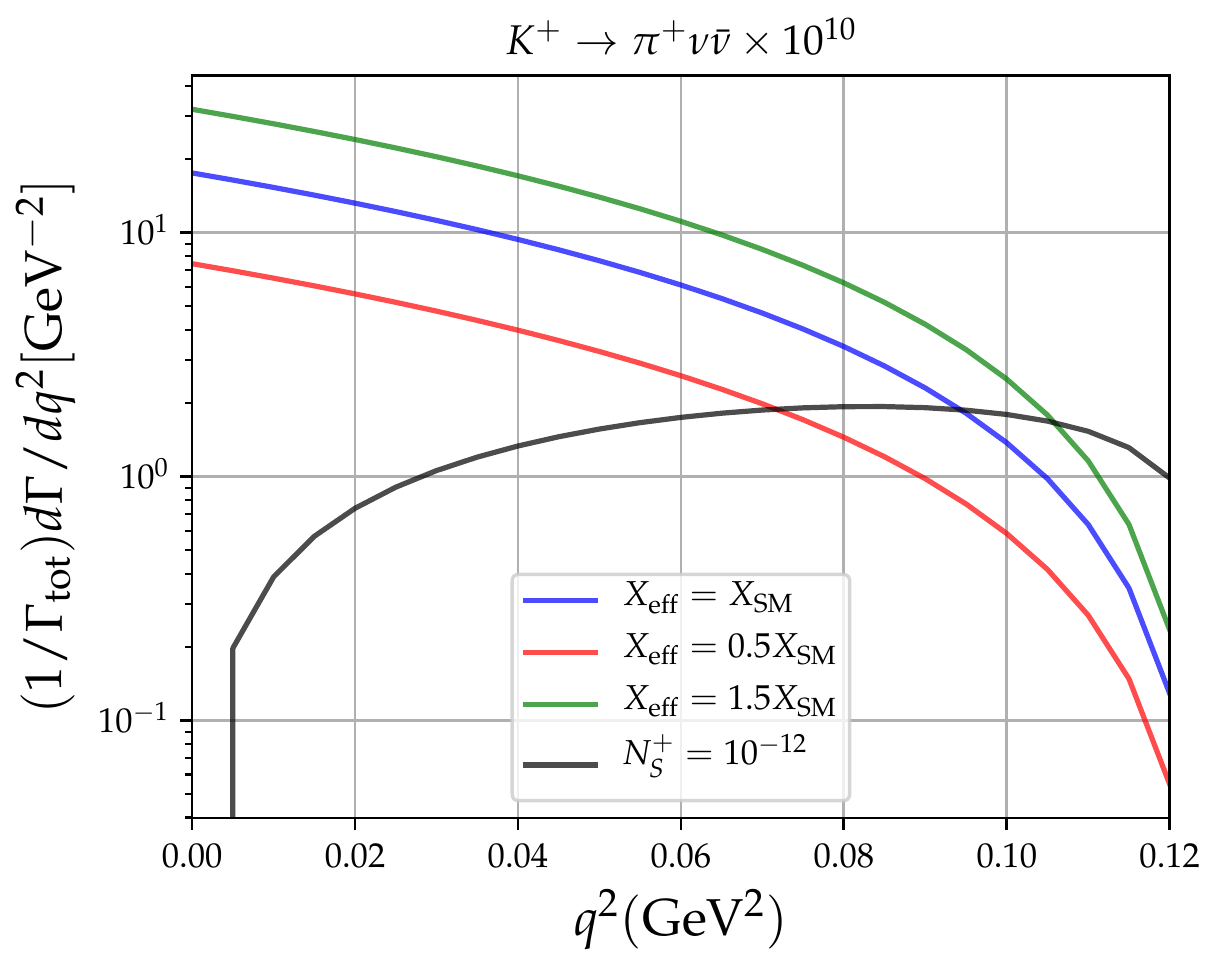}
\includegraphics[width=0.49\textwidth]{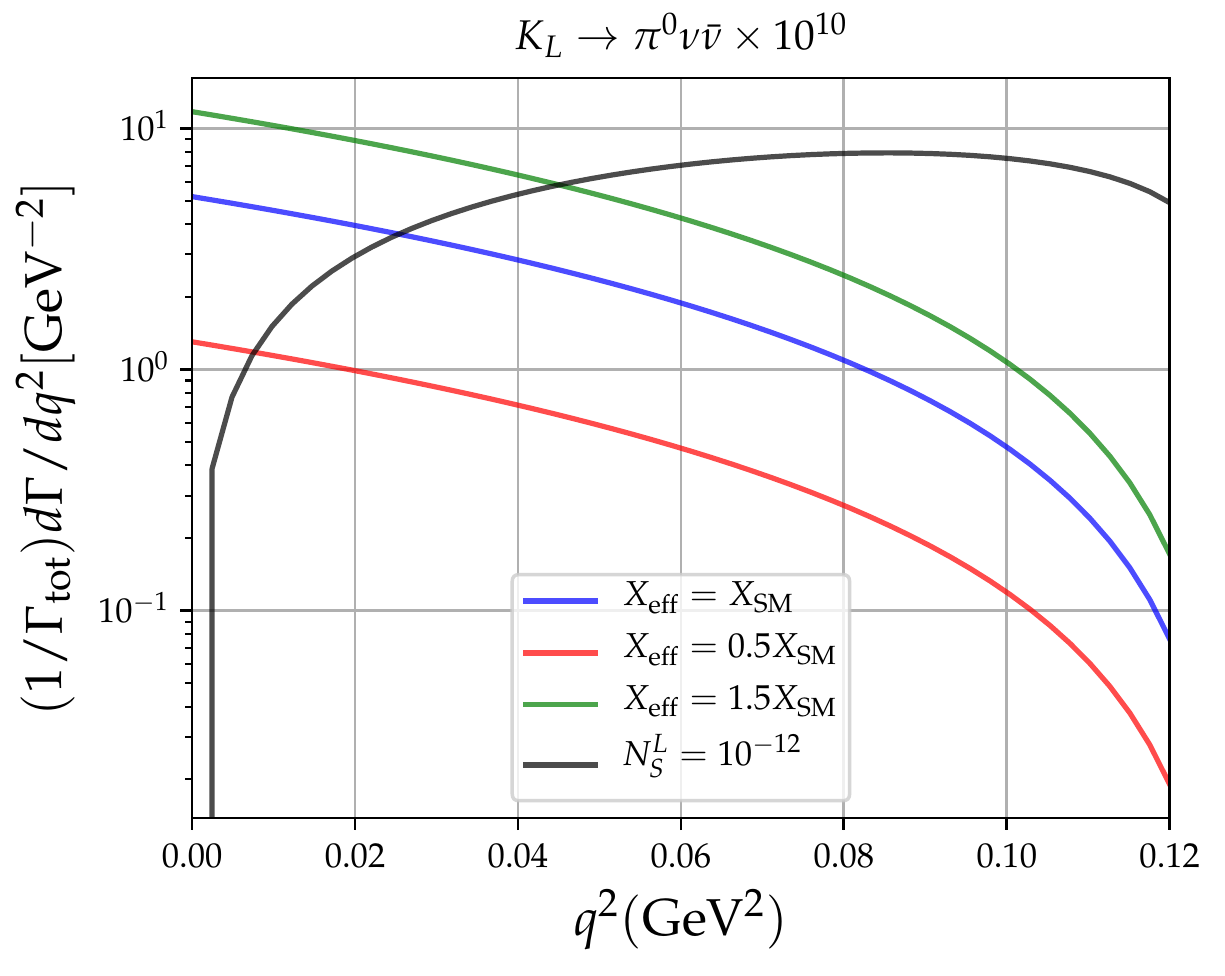}
\captionsetup{width=1\textwidth}
\caption{ The $q^2$ distributions
for $K^+ \to \pi^+ \nu \bar \nu$ (left panel) and for $K_L \to \pi^0 \nu \bar \nu$ (right panel) {normalized to the total decay rate of the corresponding Kaon} are shown.
The {scalar} contributions are given in black whereas the {vector} contributions are represented by the blue, red and green lines for {$X_{\text{eff}}=X_{\text{SM}},~ 0.5 X_{\text{SM}},~ 1.5 X_{\text{SM}}$ respectively, where $X_{\text{SM}}  =  1.462 V_{ts}^* V_{td} $.}
The factors $N_S^+$ and $N_S^L$ have been chose to be $10^{-12}$, such that the scalar and
vector contributions are of comparable size. {Furthermore, the results are scaled by a factor of $10^{10}$ for better readability.}}
\label{fig:distnunu}
\end{figure}
The plots in Fig.~\ref{fig:distnunu} demonstrate in an impressive manner
the usefulness of $q^2$ distributions in the distinctions between vector and scalar contributions. More details can be found in
\cite{Li:2019fhz,Deppisch:2020oyx} and in particular in \cite{Buras:2022hws}.

\boldmath
\subsection{$K_L\to\pi^0\ell^+\ell^-$}
\unboldmath

The differential $z=q^2/m_K^2$ distribution for these decays, taking into account only
potentially interesting terms, can be written as \cite{Isidori:2004rb}
\begin{equation}
\frac{d\Gamma}{dz}=\frac{d\Gamma_{\mathrm{CPC}}}{dz}+\frac{d\Gamma
_{\mathrm{CPV}}}{dz}~,
\end{equation}
where
\begin{align}
\frac{d\Gamma_{\mathrm{CPC}}}{dz}  &  =\frac{\alpha^{2}G_{F}^{2}m_{K}^{5}%
\beta_{\pi}(z)\beta_{\ell}(z)}{2^{11}\pi^{5}}\left\{  r_{\ell}^{2}\beta_{\ell
}^{2}(z)z|S_{0}(z)|^{2}\right\}  ,\qquad\\
\frac{d\Gamma_{\mathrm{CPV}}}{dz}  &  =\frac{\alpha^{2}G_{F}^{2}m_{K}^{5}%
\beta_{\pi}(z)\beta_{\ell}(z)}{2^{11}\pi^{5}}\left\{  r_{\ell}^{2}%
z|P_{0}(z)|^{2}+\frac{2}{3}\beta_{\pi}^{2}(z)\left(  1+\frac{2r_{\ell}^{2}}%
{z}\right)  |V_{0}(z)|^{2}\right. \nonumber\\
&  \qquad\qquad+\left[  \frac{2}{3}\beta_{\pi}^{2}(z)\left(  1+\frac{2r_{\ell
}^{2}}{z}\right)  +4r_{\ell}^{2}(2+2r_{\pi}^{2}-z)\right]  |A_{0}%
(z)|^{2}\nonumber\\
&  \left.  \qquad\qquad+4r_{\ell}^{2}(1-r_{\pi}^{2})\mathrm{Re}\left[
A_{0}(z)^{\ast}P_{0}(z)\right]  \frac{{}}{{}}\right\}  ~. \label{eq:Gamma_CPV}%
\end{align}
Here
\begin{equation}
r_{i}=\frac{m_{i}}{m_{K}}~,\quad\beta_{\ell}(z)=\left(  1-\frac{4r_{\ell}^{2}%
}{z}\right)  ^{1/2},\quad\beta_{\pi}(z)=\left(  1+r_{\pi}^{4}+z^{2}%
-2z-2r_{\pi}^{2}-2zr_{\pi}^{2}\right)  ^{1/2}\,,
\end{equation}
and the kinematical range for $z$ reads
\begin{equation}
4r_{\ell}^{2}\leq z\leq(1-r_{\pi})^{2}~.
\end{equation}

Moreover one has  the following
CPV distribution
\bea
\frac{d A_{\mathrm{FB}}}{dz}  &=& \frac{\alpha^{2}G_{F}^{2}m_{K}^{5}\beta_{\pi}%
^{2}(z)\beta_{\ell}(z)}{2^{12}\pi^{5}}~\mathrm{Re}\left[  V_{0}(z)^{\ast}\left(  4r_{\ell}^{2}\beta_{\ell
}(z)S_{0}(z)\right)  \right]~.
\eea
With a proper normalization $A_{\mathrm{FB}}$
can be identified with the forward-backward
or the energy asymmetry of the two leptons. The formfactors $S_0$, $P_0$, $V_0$ and $A_0$ can be found in \cite{Isidori:2004rb}.
A detailed numerical analysis of these formulae has been presented
  in \cite{Isidori:2004rb} and in particular in \cite{Mescia:2006jd} and
  we will not repeat it here.

{\boldmath
\section{Searching for New Physics with Rare $K$ Decays}\label{sec:3}
}
While precise measurements of all the channels discussed by us constitute important tests
of the SM, their particular role in the coming years will be in the context
of the search for NP through deviations from SM predictions.
An {extendable}
list of possible avenues for the exploration of these decays is as follows:
\vspace{0.3cm}
\begin{itemize}

\item
  To study correlations among these decays with constraints from
  $\varepsilon_K$, as already investigated
  in the past in particular in \cite{Buchalla:1994tr,Blanke:2009pq,Buras:2015yca} but also with $\epe$ and $\Delta M_K$ investigated recently
  \cite{Buras:2015jaq,Aebischer:2020mkv}. Not only an efficient search for
  new CP-violating phases can be made in this manner but also {getting a handle on}
  right-handed currents is possible.
\item
  Of importance are also correlations between these decays and rare $B$-meson
  decays, in particular with $B\to K(K^*)\nu\bar\nu$
  \cite{Buras:2001af,Altmannshofer:2009ma,Buras:2014fpa,Bause:2021cna} but also with
  $B_{s,d}\to\mu^+\mu^-$.
\item
  A classification of correlations between rare $K$ decays and rare $B$
  decays as well as quark mixing and $\epe$ can be found in Chapter 19 of
  \cite{Buras:2020xsm}. See also DNA-charts in \cite{Buras:2013ooa}, Table
  10 in \cite{Buras:2012jb} in the context of $Z^\prime$ models and
  models with induced FCNCs mediated by $Z$ as well as Table~1 in \cite{Blanke:2008yr} in the case of Randall-Sundrum models.
\item
  These decays have been analyzed in numerous SM extensions.
 Among the analyses performed in the  previous decade of particular interest are selected models with $Z'$ exchanges  \cite{Buras:2012dp,Buras:2012jb,Aebischer:2019blw}, induced FCNCs mediated by {the} $Z$ boson \cite{Bobeth:2017xry,Endo:2018gdn}
{as well as} the analyses
  of these decays in the context of lepton flavour
  universality violation (LFUV). The latter ones  are within models with vector-like
  quarks \cite{Bobeth:2016llm}, leptoquark models \cite{Bobeth:2017ecx,Fajfer:2018bfj,Marzocca:2021miv}
  and also in $K\to\pi\nu\bar\nu$ decays \cite{Bordone:2017lsy} through the presence of the
  3rd generation {neutrinos}. The tests of LFUV can also be made
  through  $K\to\pi\ell\ell'$ and $K\to\ell\ell'$ \cite{Crivellin:2016vjc}.
\item
  An important issue is the pattern of correlations within the SMEFT that
  are influenced by top Yukawa couplings. These effects have been left out
  in many papers in the past. We will return to this important issue in
  Section~\ref{sec:4}.
\item
Finally the presence of lepton number violating interactions in
$K\to\pi\nu\nu$ decays would signal the Majorana character of neutrinos.
As mentioned already above  and analysed  first in \cite{Li:2019fhz,Deppisch:2020oyx} and very recently in
\cite{Buras:2022hws}, the cases of Dirac and Majorana neutrinos
  can be distinguished  through kinematical
  distributions. They are sensitive to scalar currents representing Majorana
  neutrinos and to NP scales as high as $20\tev$. Although neutrino-less double $\beta$ decay probes
higher scales, it is limited to first generations of leptons and quarks, while the rare Kaon decays in question open up a window to
different quark and neutrino flavours.
\end{itemize}

A selected list of analyses of rare $K$ decays in specific models can be found in  Table~\ref{eprimeanomaly}. Once the experimental data improve it will be interesting to look  at these papers again possibly updating the parameters and
including new bounds from collider data at the LHC which should also improve in this decade. In this context one should also mention  the analysis in
\cite{Buras:2014zga} in which it has been demonstrated that rare $K$ decays
could provide some information about scales as short as $10^{-21}\text{m}$,
corresponding to scales in the ballpark of $100\tev$. While even much higher
scales can be probed by $K^0-\bar K^0$ mixing, rare $K$ decays, considered
simultaneously can give us a better insight into the Dirac structure of
new interactions.

  \begin{table}
\renewcommand{\arraystretch}{1.3}
\centering
\resizebox{\columnwidth}{!}{
\begin{tabular}{|c|c|c|}
\hline
  NP Scenario & References  & Decays
\\
\hline\hline
  $Z$-FCNC
& \cite{Buras:2015jaq, Bobeth:2017xry, Endo:2016tnu,Endo:2018gdn,Aebischer:2020mkv}
& $\kpn$, $\klpn$, $\epe$
\\
  $Z^\prime$
&  \cite{Buras:2012dp,Buras:2012jb,Buras:2015jaq,Aebischer:2019blw,Aebischer:2020mkv},
& $\kpn$, $\klpn$, $\Delta M_K$, $\epe$
\\
  Simplified Models
& \cite{Buras:2015yca}
  & $\klpn$, $\epe$
\\
  LHT
& \cite{Blanke:2006eb,Buras:2006wk,Blanke:2015wba}
& All $K$ decays
\\
  331 Models
& \cite{Buras:2012dp}
& Small effects in $K\to\pi\nu\bar\nu$
\\
  Vector-Like Quarks
& \cite{Bobeth:2016llm}
& $\kpn$, $\klpn$ and $\Delta M_K$
\\
  Supersymmetry
&  \cite{Buras:2004qb,Isidori:2006qy,Blazek:2014qda,Altmannshofer:2009ne}, \cite{Tanimoto:2016yfy, Kitahara:2016otd, Endo:2016aws, Crivellin:2017gks, Endo:2017ums}
& $\kpn$ and $\klpn$
\\
  2HDM
& \cite{Chen:2018ytc, Chen:2018vog}
& $\kpn$ and $\klpn$
  \\
  Universal Extra Dimensions &  \cite{Buras:2002ej,Buras:2003mk} & $\kpn$ and $\klpn$

  \\
  Randall-Sundrum models &
  \cite{Blanke:2008zb,Blanke:2008yr,Albrecht:2009xr,Casagrande:2008hr,Bauer:2009cf} & All rare $K$ decays
  \\
  Leptoquarks
& \cite{Bobeth:2017ecx,Fajfer:2018bfj,Marzocca:2021miv}
& all rare $K$ decays
\\
  SMEFT
& \cite{Aebischer:2018csl,Aebischer:2020mkv}
& several processes in $K$ and $B$ system
\\
  $\text{SU(8)}$
& \cite{Matsuzaki:2018jui}
&  $b\to s\ell^+\ell^-$, $\kpn$, $\klpn$
\\
  Diquarks
& \cite{Chen:2018dfc, Chen:2018stt}
& $\varepsilon_K$, $\kpn$, $\klpn$
\\
  Vectorlike compositeness
& \cite{Matsuzaki:2019clv}
&  $R(K^{(*)})$, $R(D^{(*)})$, $\varepsilon_K$, $\kpn$, $\klpn$
\\
\hline
\end{tabular}
}
\caption{Papers studying rare decays including other processes.
  \label{eprimeanomaly}
}
\end{table}

\section{The Impact of BSM Physics}\label{sec:4}
In \cite{Aebischer:2020mkv} a SMEFT analysis involving the rare Kaon decays $\kpn$, $\klpn$, $K_L \to \pi^0 \ell^+\ell^-$, $K_S\to\mu^+\mu^-$ and their correlation with the observables $\Delta M_K$, $\epsK$ and in particular with the ratio $\epe$ have been analysed in the context of a $Z^\prime$ and a flavour-violating $Z$ model.
In this section we review the findings of this comprehensive BSM study and present an update, taking into account the most recent experimental and theoretical results. While this study involved specific models, it illustrates well the
  general structure of correlations between various $K$ physics observables.
\subsection{SMEFT}
In this subsection we briefly review the concepts of the SMEFT, which nowadays is one of the most common ways to describe NP effects. It is governed by the following Lagrangian:

\begin{equation}
  \mathcal{L}_{\text{SMEFT}} = \mathcal{L}_{\text{SM}}^{(4)}+\sum_{d=5}^{\infty}  \sum_i\frac{C^{(d)}_i}{\Lambda^{d-4}}O_i^{(d)}\,,
\end{equation}
which contains the four-dimensional SM Lagrangian, $\mathcal{L}_{\text{SM}}^{(4)}$, as well as higher dimensional operators $O_i^{(d\geq 5)}$, weighted by their Wilson coefficients and suppressed by the NP scale $\Lambda$, at which NP effects are expected to become relevant. A complete set of non-redundant SMEFT operators up to $d=6$ has first been presented in \cite{Grzadkowski:2010es}.
BSM contributions are now parameterized in terms of the Wilson coefficients $C^{(d)}_i$, which depend on the particular NP model. The renormalization group (RG) evolution of the SMEFT Wilson coefficients from the NP scale $\Lambda$ down to the EW scale $\mu_W$ is known at the one-loop level \cite{Jenkins:2013zja,Jenkins:2013wua,Alonso:2013hga}. At $\mu_W$ the SMEFT is commonly matched onto the Weak Effective Theory (WET), by integrating out the $W,\,Z$ and Higgs boson as well as the top quark. This matching is known at the tree-level \cite{Jenkins:2017jig} and one-loop \cite{Aebischer:2015fzz,Dekens:2019ept}. The QCD and QED RG evolution of the WET at one-loop  for the complete WET Lagrangian has been known already   for some time \cite{Aebischer:2017gaw,Jenkins:2017dyc}. Recently the QCD RG evolution in WET has been extended to NLO for $\Delta F=2$ and $\Delta F=1$ non-leptonic decays in \cite{Aebischer:2020dsw} and
  \cite{Aebischer:2021raf,Aebischer:2021hws}, respectively. The QCD evolution in the SMEFT
  for $\Delta F=1$ decays are known only at the LO but for non-leptonic $\Delta F=2$ transitions an NLO analysis has just been completed \cite{Aebischer:2022hws}.

  There are several public codes on the market which deal with one or several aspects of the SMEFT and WET. A recent review can be found in \cite{Proceedings:2019rnh}. In the following we will make use of the matchrunner Python package \texttt{wilson} \cite{Aebischer:2018bkb}, as well as the Mathematica package \texttt{DsixTools} \cite{Celis:2017hod,Fuentes-Martin:2020zaz}. Finally, we will adopt the \texttt{WCxf} \cite{Aebischer:2017ugx} convention for the Wilson coefficients.

\subsection{Setup}
As a first step, we will allow for NP contributions to the observables $\epe$ and $\epsK$, writing
\be\label{GENERAL}
\frac{\varepsilon'}{\varepsilon}=\left(\frac{\varepsilon'}{\varepsilon}\right)^{\rm SM}+\left(\frac{\varepsilon'}{\varepsilon}\right)^{\rm BSM}\,, \qquad \varepsilon\equiv\varepsilon_K=
e^{i\phi_\eps}\,
\left[\varepsilon_K^{\rm SM}+\varepsilon^{\rm BSM}_K\right] \,.
\ee

A non-vanishing BSM contribution to $\epe$ is motivated by the fact that the
  SM prediction for this ratio is presently very uncertain, which is mostly due to hadronic
  uncertainties. As summarized in \cite{Buras:2020wyv}, taking the present
  estimates from the LQCD RBC-UKQCD collaboration \cite{RBC:2020kdj}, ChPT
  \cite{Cirigliano:2019ani} and Dual QCD (DQCD) \cite{Buras:2015xba,Buras:2020pjp} into account  together with isospin breaking effects, that are included
  only in ChPT and DQCD, a rather broad range
     \be
    \label{SM}
     3\times 10^{-4}\le (\epe)_\text{SM} \le 18\times 10^{-4}\,
     \ee
is still allowed. Compared with
 the   experimental world average from the NA48 \cite{Batley:2002gn} and
  KTeV \cite{AlaviHarati:2002ye, Worcester:2009qt} collaborations
  \be
    \label{EXP}
    (\epe)_\text{exp}
    = (16.6 \pm 2.3) \times 10^{-4}\,,
  \ee
{one is then  motivated to parameterize BSM contributions to $\epe$ as follows
  \cite{Buras:2015jaq}
\be\label{deltaeps}
\left(\frac{\varepsilon'}{\varepsilon}\right)^{\rm BSM}= \kepe\cdot 10^{-3}\,, \qquad   0 \le \kepe \le 1.5\,.
\ee
 We will also allow for some modest NP contribution to $\varepsilon_K$,
\be \label{DES}
(\varepsilon_K)^{\rm BSM}= \keps\cdot 10^{-3}\,, \qquad -0.2\le \keps \le 0.2 \,.
\ee
The range of $\keps$ is consistent with \cite{Bona:2006ah,Charles:2015gya,Brod:2019rzc}, but depends on whether inclusive or exclusive determinations of the CKM elements $\vub$ and $\vcb$ are used.

Anticipating then an $\epe$ anomaly as hinted within the DQCD approach
 \cite{Buras:2015xba,Buras:2020pjp},  our main goal will be  to present its implications for rare $K$ decays within the SMEFT framework. To this end we will proceed as follows.
In view of the large uncertainty of $\kepe$ several SM
parameters will be set to their central values. We choose for instance for the CKM factors and the CKM phase $\delta$:

\be\label{CKMfixed}
 {\rm Re}\lambda_t=-3.3\cdot 10^{-4}, \qquad {\rm Im}\lambda_t=1.40\cdot 10^{-4}\,, \qquad \delta= 1.15\,,
\ee
being in good agreement with estimates obtained by the UTfit \cite{Bona:2006ah} and CKMfitter \cite{Charles:2015gya} collaborations.

{\boldmath
\subsection{$Z^\prime$: A case study}}
Next, in order to illustrate NP effects in a concrete NP scenario,
in addition to the SM field content we will assume a new heavy $Z'$ boson which is governed by the following interactions with the SM fermions:
\begin{align}\label{eq:Zplag}
  \mathcal{L}_{Z'}=& -g_q^{ij} (\bar q^i \gamma^\mu q^j)Z'_{\mu}-g_u^{ij} (\bar u^i \gamma^\mu u^j)Z'_{\mu}-g_d^{ij} (\bar d^i \gamma^\mu d^j)Z'_{\mu}  \\\notag
  &-g_\ell^{ij} (\bar \ell^i \gamma^\mu \ell^j)Z'_{\mu}-g_e^{ij} (\bar e^i \gamma^\mu e^j)Z'_{\mu}\,\,+\,\, \text{h.c.}\,.
\end{align}
Having this setup at hand, we will study $\epe$, rare $K$ decays and $\Delta M_K$ in three different scenarios of the heavy $Z'$ boson.
The left-handed scenario (LHS), in which we allow for a flavour violating coupling to the left-handed fermions, the right-handed scenario (RHS), in which flavour violation results from coupling to right-handed fermions, and thirdly the left-right scenario (LR), being a mixture of the first two scenarios. The non-zero $Z'$ couplings to the SM fermions in these three cases read:
\begin{align}\label{eq:LHS}
\text{LHS}:& \quad g_q^{11,21}\,,\quad g_u^{11}\,,\quad g_d^{11}\,,\quad g_l^{11}\,,\quad g_l^{22}\,, \\
\text{RHS}:& \quad g_q^{11}\,,\quad g_u^{11}\,,\quad g_d^{11,21}\,,\quad g_l^{11}\,,\quad g_l^{22}\,,\\
\text{LR}:& \quad g_q^{11,21}\,,\quad g_u^{11}\,,\quad g_d^{11,21}\,,\quad g_l^{11}\,,\quad g_l^{22}\,.\label{eq:LR}
\end{align}

Assuming these three different scenarios, the BSM contributions to the branching ratios for $\kpn$, $\klpn$,
$K_{L,S}\to\mu^+\mu^-$ and $K_L\to\pi^0\ell^+\ell^-$
 and to $\Delta M_K$ will be discussed in the next subsections. For this purpose we will introduce the following quantities:

 \begin{eqnarray}\label{eq:kappas}
 R_{\Delta M_K}    &=&  \frac{\Delta M_K^{\rm BSM}} {\Delta M_K^{exp}} \,,\quad
 R_{\nu\bar\nu}^+  =  \frac{\mathcal{B}(K^+ \to \pi^+ \nu \bar \nu)}{\mathcal{B}(K^+ \to \pi^+ \nu \bar \nu)_{\rm SM}} \,,\quad
 R_{\nu\bar\nu}^0 =  \frac{\mathcal{B}(K_L \to \pi^0 \nu \bar \nu)}{\mathcal{B}(K_L \to \pi^0 \nu \bar \nu)_{\rm SM}}\,, \\\notag
 R_{\mu^+\mu^-}^S  &=&  \frac{\mathcal{B}(K_S \to \mu^+\mu^-)}{\mathcal{B}(K_S \to \mu^+\mu^-)_{\rm SM}} \,,\quad R_{\pi\ell^+\ell^-}^0 =  \frac{\mathcal{B}(K_L \to \pi^0 \ell^+ \ell^-)}{\mathcal{B}(K_L \to \pi^0 \ell^+ \ell^-)_{\rm SM}}\,.
 \end{eqnarray}

Now we discuss the different $Z'$ scenarios defined in \eqref{eq:LHS}-\eqref{eq:LR}, namely the LHS, the RHS and the LR scenario. In the following we will assume a heavy $Z'$ boson with a mass of $M_{Z'}=10\tev$ to evade constraints from direct searches \cite{ATLAS:2019erb, CMS:2019tbu}.

\subsubsection{LHS scenario}

We start our discussion with the so called electroweak penguin (EWP) scenario defined as follows:

\begin{equation}
g_q^{21}\neq 0\,, \qquad  g_u^{11}=-2g_d^{11}\,, \qquad g_{\ell}^{11,22} \neq0 \qquad \text{(EWP scenario)}\,.
\end{equation}
This choice of parameters generates after matching and running to the EW scale the EWP operator

\begin{equation}
Q_8=6\,(\bar s^\alpha\gamma_\mu P_Ld^\beta)\sum_q Q_q (\bar q^\beta \gamma^\mu P_R \,q^\alpha)\,,
\end{equation}
with the electric charge $Q_q$ of the quark $q$.

\begin{figure}[htb]
\centering
\includegraphics[width=0.45\textwidth]{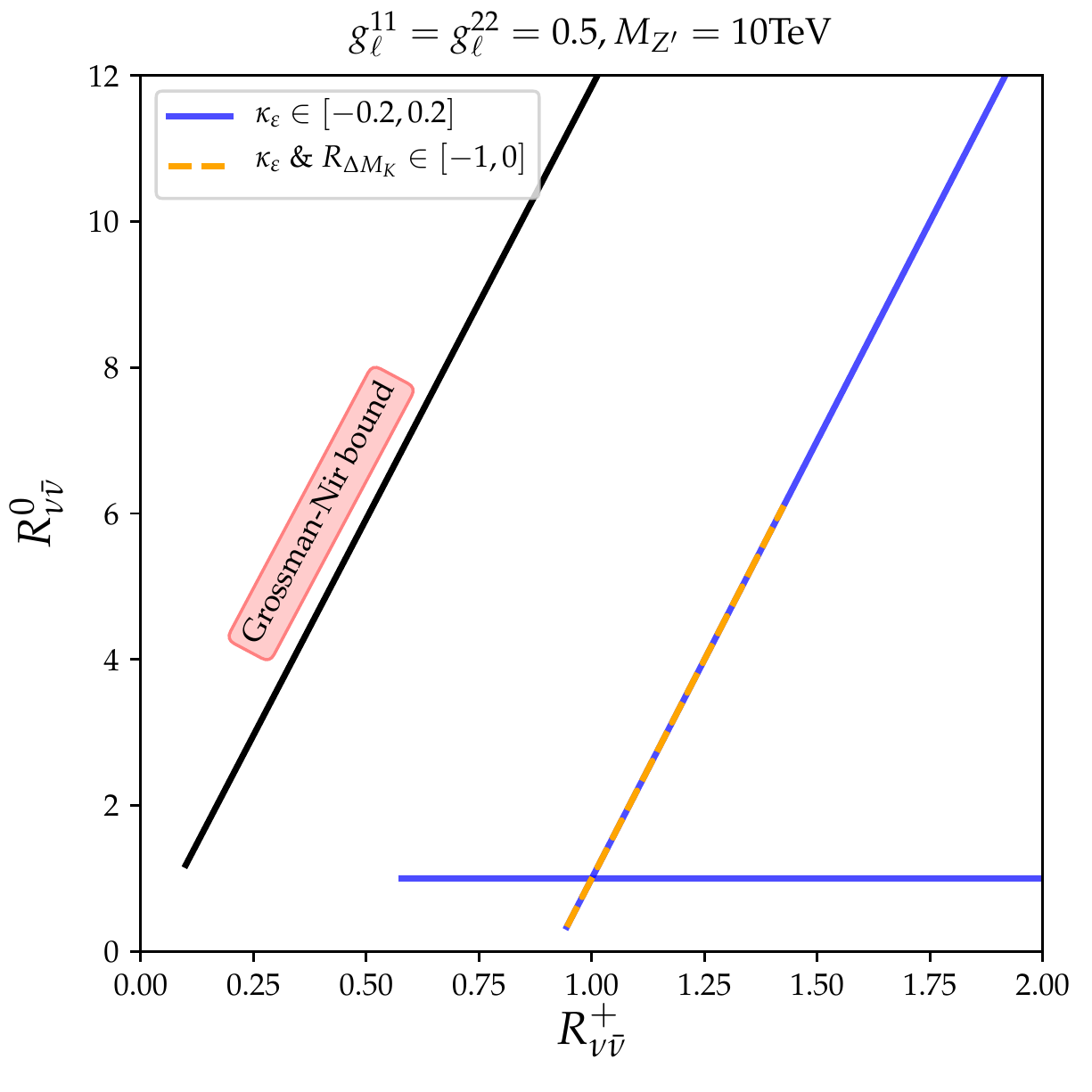}
\includegraphics[width=0.45\textwidth]{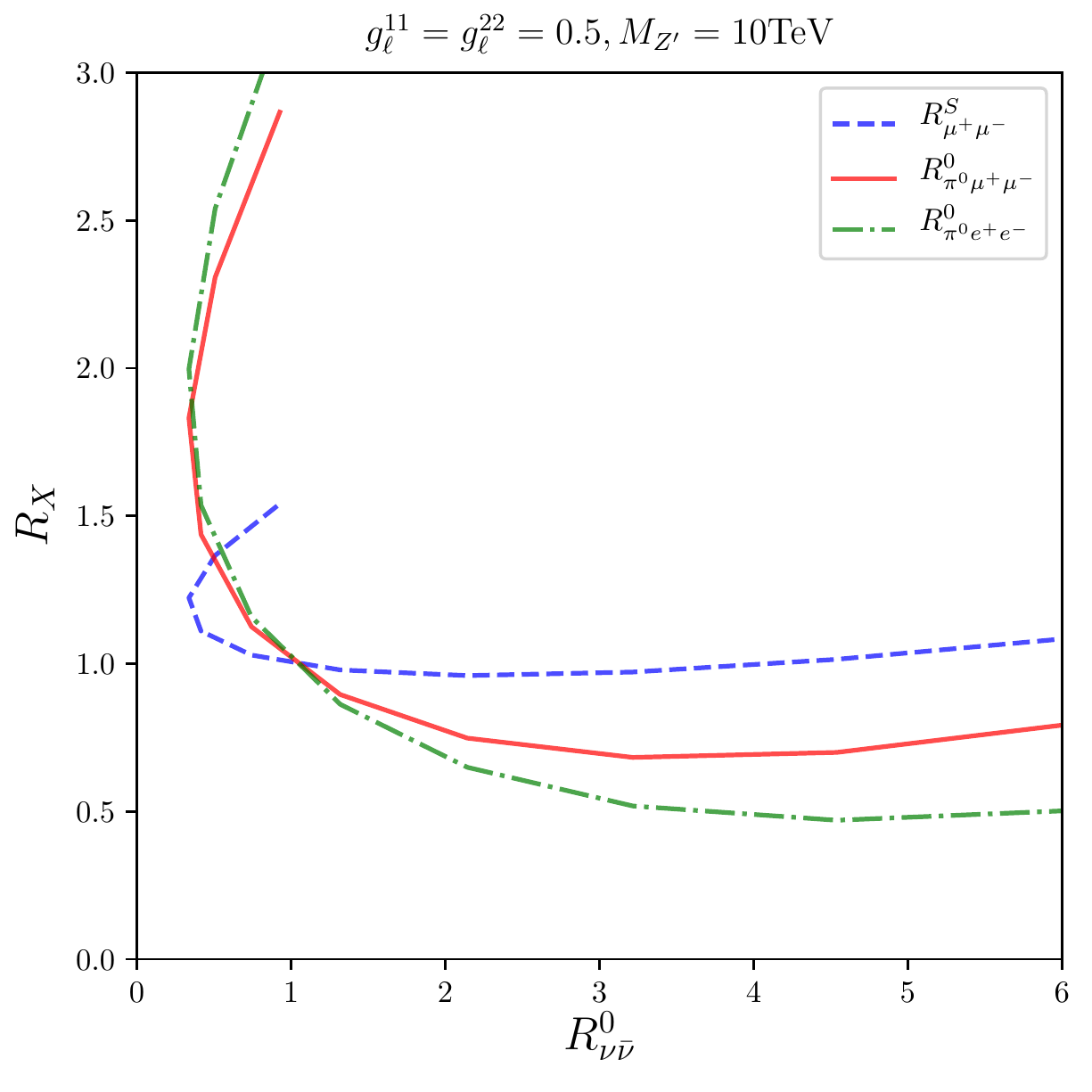}
\captionsetup{width=0.9\textwidth}
\caption{The EWP scenario for a $Z'$ of $10\tev$. The correlation between the ratios for the process $K^+\to\pi^+\nu\bar\nu$, $K_L\to\pi\nu\bar\nu$ defined in \eqref{eq:kappas}
	is plotted (left). The blue (orange) lines are allowed by $\kappa_{\varepsilon}$
	($\kappa_\varepsilon$ and $R_{\Delta M_K}$) constraints {and the black line represents the GN bound}. The correlations between the
	ratio for $K_L\to \pi^0\nu\bar\nu$ and the ones for
	$K \to\pi \ell^+ \ell^-$ and  $K_S \to \mu^+ \mu^-$ after imposing $\kappa_\varepsilon$ and $R_{\Delta M_K}$ constraints are shown in the right panel.}
\label{fig:LHS_kp_k0}
\end{figure}

\begin{figure}[htb]
\centering
\includegraphics[width=0.45\textwidth]{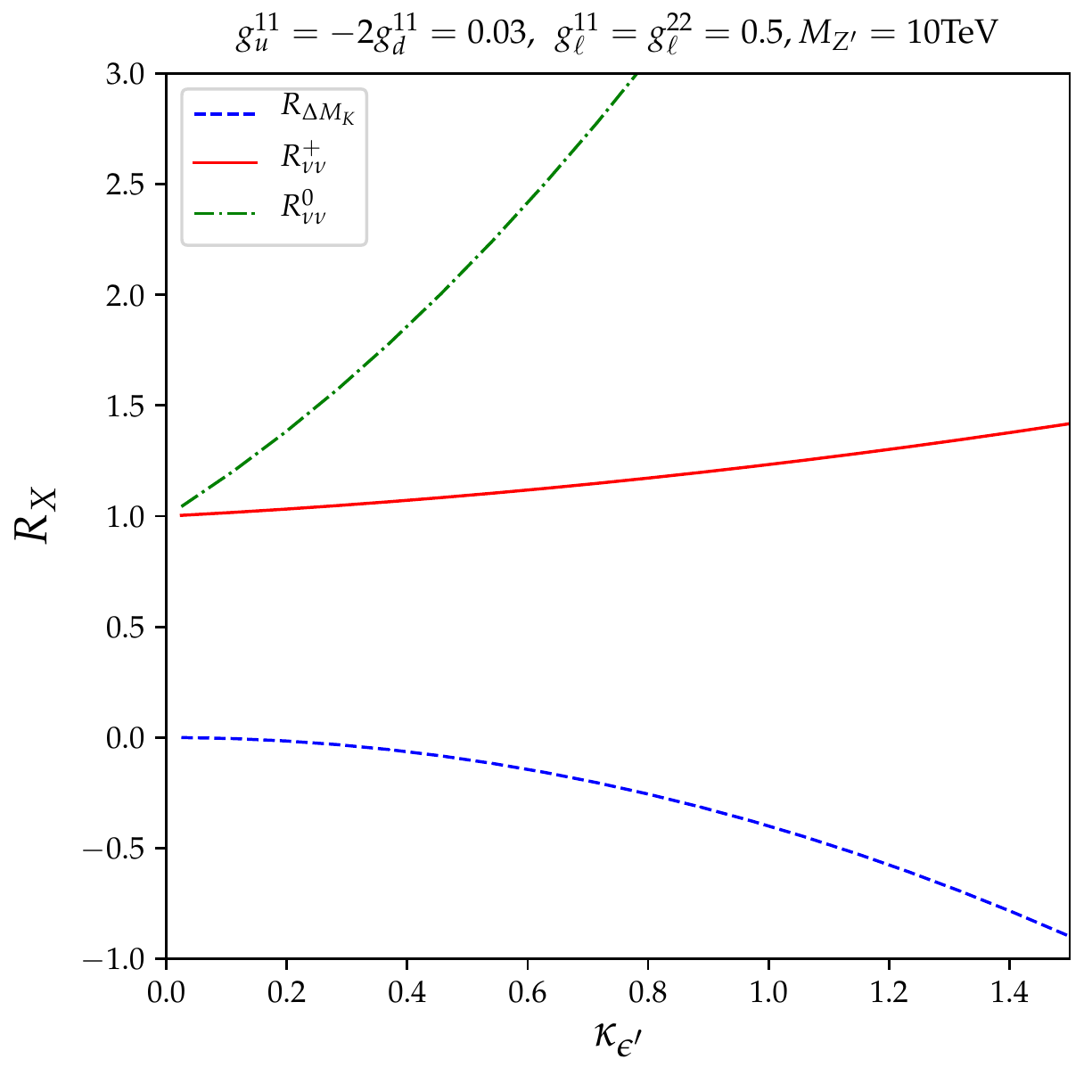}
\includegraphics[width=0.45\textwidth]{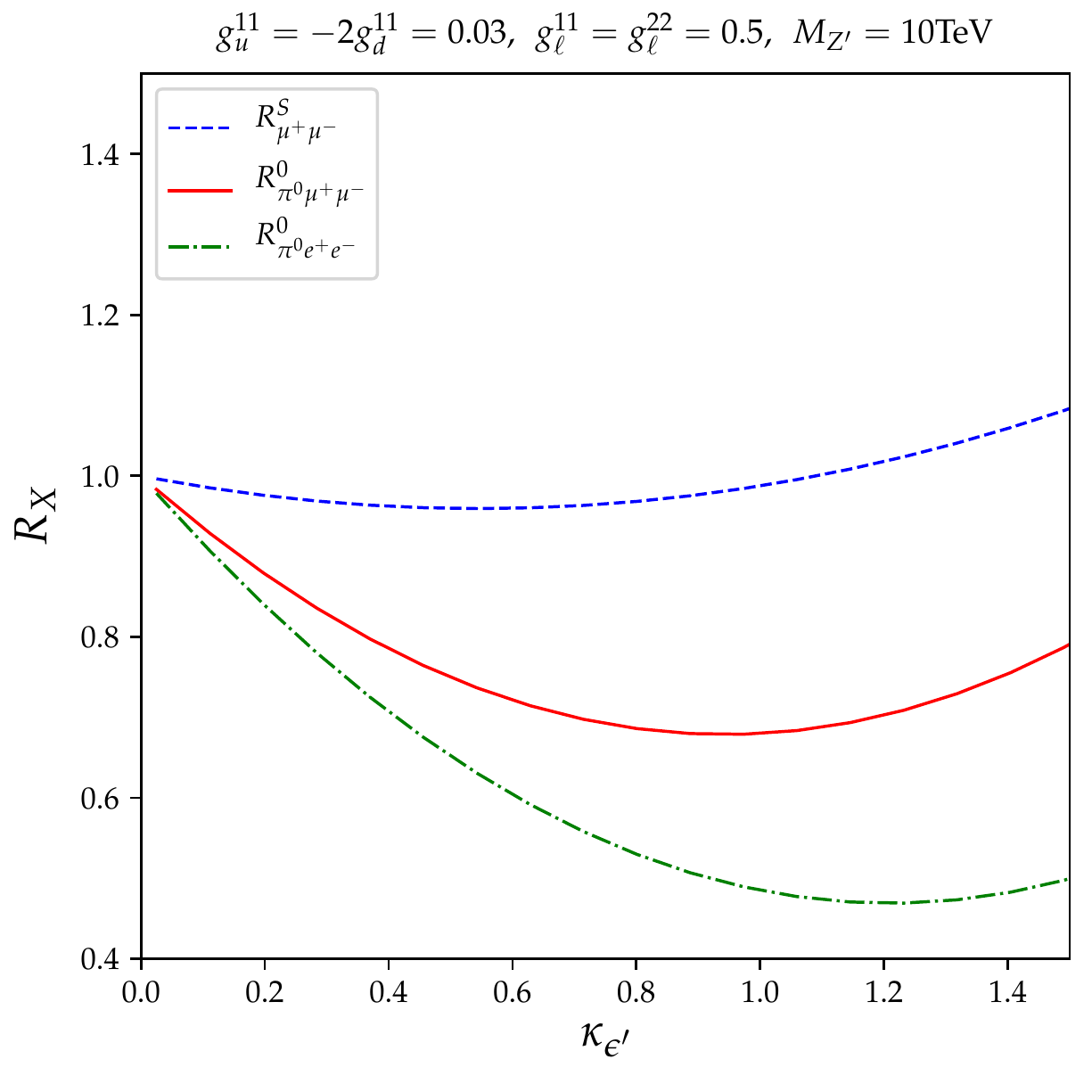}
\captionsetup{width=0.9\textwidth}
\caption{The EWP scenario for a $Z'$ of $10\tev$. The ratios for $\Delta M_K$ and for the process $K^+\to\pi^+\nu\bar\nu$, $K_L\to\pi\nu\bar\nu$ defined in \eqref{eq:kappas} are plotted against $\kappa_{\varepsilon'}$ (left panel) and the predictions for the ratios of the decays $K_S \to \mu^+ \mu^-$, $K_L \to \pi^0 \mu^+ \mu^-$ and $K_L \to \pi^0 e^+ e^-$ defined in eq.~\eqref{eq:kappas} are plotted against $\kappa_{\varepsilon'}$ (right panel).
}
\label{fig:LHS_kappas2}
\end{figure}

The results are shown in Figs. \ref{fig:LHS_kp_k0} and \ref{fig:LHS_kappas2}. They are self-explanatory but let us mention several of observations:
  \begin{itemize}
  \item
    The  $\varepsilon_K$  constraint forces  in this scenario
    the correlation between the branching ratios for
    $\kpn$ and $\klpn$ decays to take place, as seen in the left panel of Fig.~\ref{fig:LHS_kp_k0}, on the two solid blue lines \cite{Blanke:2009pq}. But
    as pointed out in  \cite{Aebischer:2020mkv} when also
    the $\Delta M_K$ constraint is taken into account the horizontal line
    is excluded.
  \item
    As seen in Fig.~\ref{fig:LHS_kappas2} the $\epe$ anomaly, if confirmed, will
    have a large impact on all decays except $\kpn$.
\end{itemize}

\subsubsection{RHS {scenario}}
In this section we discuss the right-handed QCD penguin (QCDP) scenario, defined by the following choice of parameters:
\begin{equation}
g_q^{11}\neq 0\,,\qquad  g_d^{21} \ne 0\,, \qquad \text{(RH-QCDP scenario)}
\end{equation}
which generates the QCP operator

\begin{equation}
Q_6=4\,(\bar s^\alpha\gamma_\mu P_Ld^\beta)\sum_q (\bar q^\beta \gamma^\mu P_R \,q^\alpha)\,,
\end{equation}
at the EW scale. However, this scenario is excluded due to $\kappa_{\epsilon}$,
originating from RG running effects from the NP scale down to the EW scale\cite{Aebischer:2020mkv}. Further details to this so-called back-rotation effect are discussed in \cite{Aebischer:2020lsx}.

\subsubsection{LR scenario}
Finally we discuss a combination of left-handed and right-handed flavour changing couplings, represented by the LR-EWP scenario:

\begin{equation}
g_q^{21}, g_d^{21} \neq 0\,, \qquad  g_u^{11}=-2g_d^{11}\,. \qquad \text{(LR-EWP scenario)}
\end{equation}
In such a scenario not only the two branches like in the LHS but the full parameter space can be reached, by allowing for right-handed couplings. We show a particular example in Fig.~\ref{fig:epsK}.
It shows the GN bound in black, together with the two-branch system in blue, which is also given in Fig.~\ref{fig:LHS_kp_k0} (left). Imposing the constraint from $\Delta M_K$ only allows for points on the tilted Monika-Blanke branch (orange), as discussed above and also in \cite{Blanke:2009pq}. The LR model shown in green allows to populate all parameter space, which is allowed by the GN bound. This fact is also illustrated by the red region in Fig.~\ref{fig:illustrateEpsK}.

\begin{figure}[htb]
\centering
\includegraphics[width=0.6\textwidth]{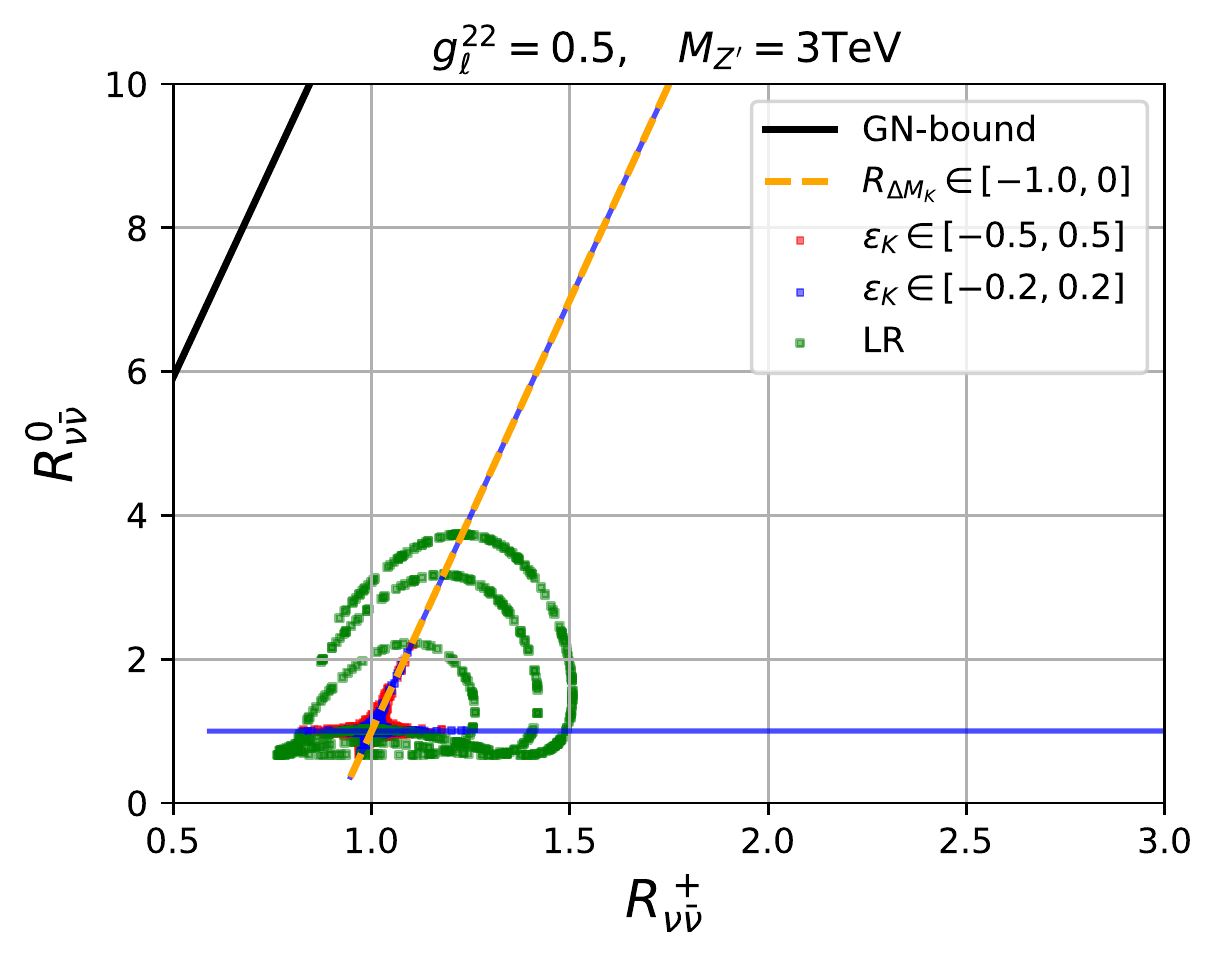}
\captionsetup{width=0.9\textwidth}
	\caption{{The $K^+\to\pi^+\nu\bar\nu$ and $K_L\to\pi\nu\bar\nu$ ratios defined in
\eqref{eq:kappas} are shown. The LR scenario is shown in green, the LH-EWP scenario in blue and red with
 $\varepsilon_K \in [-0.2,0.2]$ and $[-0.5,0.5]$ for a $Z'$ of $3\tev$. Points satisfying the $R_{\Delta M_K} \in [-1.0,0]$ constraint are shown in orange. The black line represents the GN bound.}}
\label{fig:epsK}
\end{figure}

{\boldmath
\subsection{$Z$ from $Z^\prime$}}
In this section an example is shown, where flavour violating effects result from modified $Z$-couplings through RG mixing. Choosing the following parameters:
\begin{equation}
g_q^{21}\neq 0\,,\quad	g_u^{11} = g_d^{11}= 0\,, \quad g_q^{33} = 2\,,
\end{equation}
induces through large Yukawa running effects flavour-violating $Z-s-d$ couplings. The effects of such new couplings are shown for the different observables in Fig.~\ref{fig:epsp-phiq-RGE}.

The scenario has the largest impact on the observable $\Delta M_K$, which is reduced for positive NP contributions to $\epe$. Also $\klpn$ becomes smaller for a constructive contribution to $\epe$. All the other rare Kaon decays and in particular $\kpn$ are to a good approximation not affected in this scenario.

\begin{figure}[htb]
\centering
\includegraphics[width=0.7\textwidth]{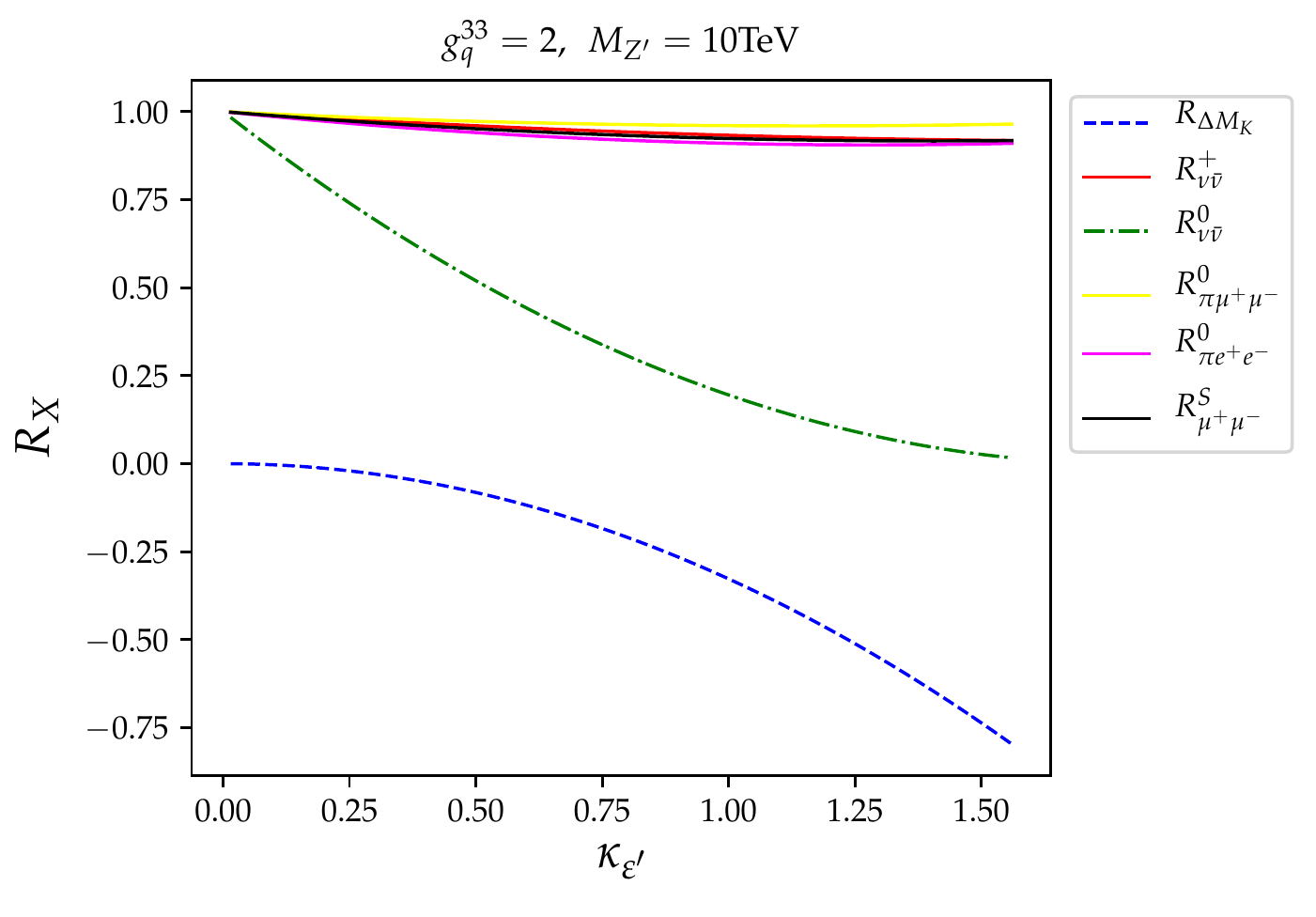}
\captionsetup{width=0.9\textwidth}
\caption{Shown are the $Z$-contributions to $\epe$ and other Kaon
observables, generated from a $Z^\prime$
 with purely left-handed quark couplings through RG running.}
\label{fig:epsp-phiq-RGE}
\end{figure}


\section{Outlook}\label{sec:5}
In our paper we have concentrated on rare $K$ decays that will be measured
in this decade at various laboratories. In the coming years
the main role will be played by the measurements of
$\kpn$ by the NA62 collaboration at CERN SPS \cite{Ceccucci:2018bnw},
 and for $\klpn$ by KOTO \cite{Ahn:2018mvc} at J-PARC and later by
 KLEVER \cite{Ambrosino:2019qvz} at CERN SPS. Already  these measurements
 will allow for a deep insight into possible NP at short
 distances, in particular if also $q^2$ distributions will be measured.
 Only in the second half of this decade we will be able to benefit
 from the measurements of $\ksm$ and $K_L\to\pi^0\ell^+\ell^-$ decays.

 But it should be emphasized, as done in particular in \cite{Buras:2020xsm,Buras:2013ooa}, that an important role in the search for NP
 is played by correlations between decays considered by us and other observables.
  Therefore this decade should be very exciting for flavour phenomenology, not only because of the decays considered here but also due to $B$ decays explored
  at Belle II \cite{Belle-II:2018jsg} and LHCb \cite{Cerri:2018ypt,Bediaga:2018lhg}.
  Moreover  also ATLAS and CMS will contribute in an important
manner  \cite{Cerri:2018ypt,CidVidal:2018eel} and generally BSM searches
beyond colliders at CERN \cite{Beacham:2019nyx}. Moreover, with the advances
in LQCD it will be possible to make clear cut conclusions about the presence of NP in processes in which hadronic effects play an important role \cite{Cerri:2018ypt,USQCD:2019hyg,Cirigliano:2019jig,Joo:2019byq}.

\section*{Acknowledgements}

We thank Gino Isidori, Julia Harz and Elena Venturini for the discussions.
J.\,A.\ acknowledges financial support from the European Research Council (ERC) under the European Union’s Horizon 2020 research and innovation programme under grant agreement 833280 (FLAY), and from the Swiss National Science Foundation (SNF) under contract 200020-204428.
A.J.B acknowledges financial support from the Excellence Cluster ORIGINS,
funded by the Deutsche Forschungsgemeinschaft (DFG, German Research Foundation)
under Germany's Excellence Strategy – EXC-2094 – 390783311.
J.K. is financially supported by the Alexander von Humboldt Foundation’s postdoctoral
research fellowship.

\addcontentsline{toc}{section}{References}

\small

\bibliographystyle{JHEP}
\bibliography{Bookallrefs}

\providecommand{\href}[2]{#2}\begingroup\raggedright\begin{thebibliography}{100}

\bibitem{Ceccucci:2018bnw}
A.~Ceccucci, {\it {Review and Outlook on Kaon Physics}},  {\em Acta Phys.
  Polon.} {\bf B49} (2018) 1079--1086.

\bibitem{Cerri:2018ypt}
A.~Cerri, V.~V. Gligorov, S.~Malvezzi, J.~Martin~Camalich, and J.~Zupan, {\it
  {Opportunities in Flavour Physics at the HL-LHC and HE-LHC}},
  \href{http://arxiv.org/abs/1812.07638}{{\tt arXiv:1812.07638}}.

\bibitem{Bediaga:2018lhg}
{\bf LHCb} Collaboration, R.~Aaij et~al., {\it {Physics case for an LHCb
  Upgrade II - Opportunities in flavour physics, and beyond, in the HL-LHC
  era}},  \href{http://arxiv.org/abs/1808.08865}{{\tt arXiv:1808.08865}}.

\bibitem{Ahn:2018mvc}
{\bf KOTO} Collaboration, J.~Ahn et~al., {\it {Search for the $K_L \!\to\!
  \pi^0 \nu \overline{\nu}$ and $K_L \!\to\! \pi^0 X^0$ decays at the J-PARC
  KOTO experiment}},  {\em Phys. Rev. Lett.} {\bf 122} (2019), no.~2 021802,
  [\href{http://arxiv.org/abs/1810.09655}{{\tt arXiv:1810.09655}}].

\bibitem{Ambrosino:2019qvz}
{\bf KLEVER Project} Collaboration, F.~Ambrosino et~al., {\it {KLEVER: An
  experiment to measure BR($K_L\to\pi^0\nu\bar{\nu}$) at the CERN SPS}},
  \href{http://arxiv.org/abs/1901.03099}{{\tt arXiv:1901.03099}}.

\bibitem{Buras:2020xsm}
A.~J. Buras, {\em {Gauge Theory of Weak Decays}}.
\newblock Cambridge University Press, 6, 2020.

\bibitem{Kiyo:1998zm}
Y.~Kiyo, T.~Morozumi, P.~Parada, M.~N. Rebelo, and M.~Tanimoto, {\it {Quark
  mass hierarchy, FCNC and CP violation in a seesaw model}},  {\em Prog. Theor.
  Phys.} {\bf 101} (1999) 671--706,
  [\href{http://arxiv.org/abs/hep-ph/9809333}{{\tt hep-ph/9809333}}].

\bibitem{Mescia:2007kn}
F.~Mescia and C.~Smith, {\it {Improved estimates of rare K decay
  matrix-elements from $K_{\ell3}$ decays}},  {\em Phys.~Rev.} {\bf D76} (2007)
  034017, [\href{http://arxiv.org/abs/0705.2025}{{\tt arXiv:0705.2025}}].

\bibitem{Buchalla:2003sj}
G.~Buchalla, G.~D'Ambrosio, and G.~Isidori, {\it {Extracting short-distance
  physics from $K_{L,S} \to \pi^0 e^+ e^-$ decays}},  {\em Nucl.~Phys.} {\bf
  B672} (2003) 387--408, [\href{http://arxiv.org/abs/hep-ph/0308008}{{\tt
  hep-ph/0308008}}].

\bibitem{Isidori:2003ts}
G.~Isidori and R.~Unterdorfer, {\it {On the short-distance constraints from
  $K_{L,S} \to \mu^+ \mu^-$}},  {\em JHEP} {\bf 01} (2004) 009,
  [\href{http://arxiv.org/abs/hep-ph/0311084}{{\tt hep-ph/0311084}}].

\bibitem{DAmbrosio:2017klp}
G.~D'Ambrosio and T.~Kitahara, {\it {Direct $CP$ Violation in $K \to \mu^+
  \mu^-$}},  {\em Phys. Rev. Lett.} {\bf 119} (2017), no.~20 201802,
  [\href{http://arxiv.org/abs/1707.06999}{{\tt arXiv:1707.06999}}].

\bibitem{Mescia:2006jd}
F.~Mescia, C.~Smith, and S.~Trine, {\it {$K_L\to\pi^0 e^+e^-$ and
  $K_L\to\pi^0\mu^+\mu^-$: A binary star on the stage of flavor physics}},
  {\em JHEP} {\bf 08} (2006) 088,
  [\href{http://arxiv.org/abs/hep-ph/0606081}{{\tt hep-ph/0606081}}].

\bibitem{DAmbrosio:2018ytt}
G.~D'Ambrosio, D.~Greynat, and M.~Knecht, {\it {On the amplitudes for the
  CP-conserving $K^\pm(K_S)\to\pi^\pm(\pi^0)\ell^+\ell^-$ rare decay modes}},
  {\em JHEP} {\bf 02} (2019) 049, [\href{http://arxiv.org/abs/1812.00735}{{\tt
  arXiv:1812.00735}}].

\bibitem{DAmbrosio:2019xph}
G.~D'Ambrosio, D.~Greynat, and M.~Knecht, {\it {Matching long and short
  distances at order ${\mathcal O}(\alpha_s)$ in the form factors for $K\to\pi
  \ell^+\ell^-$}},  {\em Phys. Lett. B} {\bf 797} (2019) 134891,
  [\href{http://arxiv.org/abs/1906.03046}{{\tt arXiv:1906.03046}}].

\bibitem{Dery:2021mct}
A.~Dery, M.~Ghosh, Y.~Grossman, and S.~Schacht, {\it {$K\to\mu^+\mu^-$ as a
  clean probe of short-distance physics}},  {\em JHEP} {\bf 07} (2021) 103,
  [\href{http://arxiv.org/abs/2104.06427}{{\tt arXiv:2104.06427}}].

\bibitem{DAmbrosio:1996kjn}
G.~D'Ambrosio and J.~Portoles, {\it {Vector meson exchange contributions to $K
  \to \pi \gamma \gamma$ and $K_L \to \gamma \ell^+ \ell^-$}},  {\em Nucl.
  Phys.} {\bf B492} (1997) 417--454,
  [\href{http://arxiv.org/abs/hep-ph/9610244}{{\tt hep-ph/9610244}}].

\bibitem{Gerard:2005yk}
J.-M. G\'erard, C.~Smith, and S.~Trine, {\it {Radiative kaon decays and the
  penguin contribution to the $\Delta I = 1/2$ rule}},  {\em Nucl.~Phys.} {\bf
  B730} (2005) 1--36, [\href{http://arxiv.org/abs/hep-ph/0508189}{{\tt
  hep-ph/0508189}}].

\bibitem{GomezDumm:1998gw}
D.~Gomez~Dumm and A.~Pich, {\it {Long distance contributions to the
  $K_L\to\mu^+\mu^-$ decay width}},  {\em Phys. Rev. Lett.} {\bf 80} (1998)
  4633--4636, [\href{http://arxiv.org/abs/hep-ph/9801298}{{\tt
  hep-ph/9801298}}].

\bibitem{Buchalla:1994tr}
G.~Buchalla and A.~J. Buras, {\it {$\sin2\beta$ from $K \to \pi \nu\bar\nu$}},
  {\em Phys.~Lett.} {\bf B333} (1994) 221--227,
  [\href{http://arxiv.org/abs/hep-ph/9405259}{{\tt hep-ph/9405259}}].

\bibitem{Bordone:2019guc}
M.~Bordone, N.~Gubernari, D.~van Dyk, and M.~Jung, {\it {Heavy-Quark expansion
  for ${{\bar{B}}_s\rightarrow D^{(*)}_s}$ form factors and unitarity bounds
  beyond the ${SU(3)_F}$ limit}},  {\em Eur. Phys. J. C} {\bf 80} (2020), no.~4
  347, [\href{http://arxiv.org/abs/1912.09335}{{\tt arXiv:1912.09335}}].

\bibitem{Bordone:2021oof}
M.~Bordone, B.~Capdevila, and P.~Gambino, {\it {Three loop calculations and
  inclusive $\vcb$}},  {\em Phys. Lett. B} {\bf 822} (2021) 136679,
  [\href{http://arxiv.org/abs/2107.00604}{{\tt arXiv:2107.00604}}].

\bibitem{Ricciardi:2021shl}
G.~Ricciardi, {\it {Semileptonic $B$ decays and $|V_{xb}|$ update}},  in {\em
  {19th International Conference on B-Physics at Frontier Machines}}, 3, 2021.
\newblock \href{http://arxiv.org/abs/2103.06099}{{\tt arXiv:2103.06099}}.

\bibitem{Aoki:2019cca}
{\bf Flavour Lattice Averaging Group} Collaboration, S.~Aoki et~al., {\it {FLAG
  Review 2019: Flavour Lattice Averaging Group (FLAG)}},  {\em Eur. Phys. J. C}
  {\bf 80} (2020), no.~2 113, [\href{http://arxiv.org/abs/1902.08191}{{\tt
  arXiv:1902.08191}}].

\bibitem{Buras:2021nns}
A.~J. Buras and E.~Venturini, {\it {Searching for New Physics in Rare $K$ and
  $B$ Decays without $|V_{cb}|$ and $|V_{ub}|$ Uncertainties}},
  \href{http://arxiv.org/abs/2109.11032}{{\tt arXiv:2109.11032}}.

\bibitem{Gaillard:1974hs}
M.~Gaillard and B.~W. Lee, {\it {Rare Decay Modes of the K-Mesons in Gauge
  Theories}},  {\em Phys.~Rev.} {\bf D10} (1974) 897.

\bibitem{Inami:1980fz}
T.~Inami and C.~Lim, {\it {Effects of Superheavy Quarks and Leptons in
  Low-Energy Weak Processes $K_L\to\mu^+\mu^-$, $K^+\to\pi^+\nu\bar\nu$ and
  $K^0-\bar K^0$}},  {\em Prog.~Theor.~Phys.} {\bf 65} (1981) 297.

\bibitem{Ellis:1982ve}
J.~R. Ellis and J.~S. Hagelin, {\it {Constraints on Light Particles from Kaon
  Decays}},  {\em Nucl. Phys.} {\bf B217} (1983) 189--214.

\bibitem{Dib:1989cc}
C.~Dib, I.~Dunietz, and F.~J. Gilman, {\it {Strong Interaction Corrections to
  the Decay $K \to \pi$ Neutrino Anti-neutrino for Large M(t)}},  {\em Mod.
  Phys. Lett.} {\bf A6} (1991) 3573--3582.

\bibitem{Buchalla:1993bv}
G.~Buchalla and A.~J. Buras, {\it {QCD corrections to rare $K$ and $B$ decays
  for arbitrary top quark mass}},  {\em Nucl.~Phys.} {\bf B400} (1993)
  225--239.

\bibitem{Buchalla:1993wq}
G.~Buchalla and A.~J. Buras, {\it {The rare decays $K^+ \to \pi^+\nu\bar\nu$
  and $K_L\to\mu^+ \mu^-$ beyond leading logarithms}},  {\em Nucl.~Phys.} {\bf
  B412} (1994) 106--142, [\href{http://arxiv.org/abs/hep-ph/9308272}{{\tt
  hep-ph/9308272}}].

\bibitem{Buras:1994qa}
A.~J. Buras, M.~E. Lautenbacher, M.~Misiak, and M.~Munz, {\it {Direct CP
  violation in $K_L\to \pi^0 e^+ e^-$ beyond leading logarithms}},  {\em Nucl.
  Phys.} {\bf B423} (1994) 349--383,
  [\href{http://arxiv.org/abs/hep-ph/9402347}{{\tt hep-ph/9402347}}].

\bibitem{Misiak:1999yg}
M.~Misiak and J.~Urban, {\it {QCD corrections to FCNC decays mediated by Z
  penguins and W boxes}},  {\em Phys.~Lett.} {\bf B451} (1999) 161--169,
  [\href{http://arxiv.org/abs/hep-ph/9901278}{{\tt hep-ph/9901278}}].

\bibitem{Buchalla:1998ba}
G.~Buchalla and A.~J. Buras, {\it {The rare decays $K\to\pi \nu\bar\nu$, $B\to
  X \nu\bar\nu$ and $B\to \ell^+\ell^-$: An Update}},  {\em Nucl.~Phys.} {\bf
  B548} (1999) 309--327, [\href{http://arxiv.org/abs/hep-ph/9901288}{{\tt
  hep-ph/9901288}}].

\bibitem{Buras:2005gr}
A.~J. Buras, M.~Gorbahn, U.~Haisch, and U.~Nierste, {\it {The rare decay $K^+
  \to \pi^+ \nu \bar\nu$ at the next-to-next-to-leading order in QCD}},  {\em
  Phys.~Rev.~Lett.} {\bf 95} (2005) 261805,
  [\href{http://arxiv.org/abs/hep-ph/0508165}{{\tt hep-ph/0508165}}].

\bibitem{Buras:2006gb}
A.~J. Buras, M.~Gorbahn, U.~Haisch, and U.~Nierste, {\it {Charm quark
  contribution to $K^+ \to \pi^+ \nu \bar\nu$ at next-to-next-to-leading
  order}},  {\em JHEP} {\bf 11} (2006) 002,
  [\href{http://arxiv.org/abs/hep-ph/0603079}{{\tt hep-ph/0603079}}].

\bibitem{Gorbahn:2004my}
M.~Gorbahn and U.~Haisch, {\it {Effective Hamiltonian for non-leptonic $|\Delta
  F| = 1$ decays at NNLO in QCD}},  {\em Nucl.~Phys.} {\bf B713} (2005)
  291--332, [\href{http://arxiv.org/abs/hep-ph/0411071}{{\tt hep-ph/0411071}}].

\bibitem{Isidori:2005xm}
G.~Isidori, F.~Mescia, and C.~Smith, {\it {Light-quark loops in $K
  \to\pi\nu\bar\nu$}},  {\em Nucl.~Phys.} {\bf B718} (2005) 319--338,
  [\href{http://arxiv.org/abs/hep-ph/0503107}{{\tt hep-ph/0503107}}].

\bibitem{Brod:2008ss}
J.~Brod and M.~Gorbahn, {\it {Electroweak Corrections to the Charm Quark
  Contribution to $K^+ \to \pi^+ \nu \bar\nu$}},  {\em Phys.~Rev.} {\bf D78}
  (2008) 034006, [\href{http://arxiv.org/abs/0805.4119}{{\tt
  arXiv:0805.4119}}].

\bibitem{Brod:2010hi}
J.~Brod, M.~Gorbahn, and E.~Stamou, {\it {Two-Loop Electroweak Corrections for
  the $K \to \pi \nu \bar{\nu}$ Decays}},  {\em Phys.~Rev.} {\bf D83} (2011)
  034030, [\href{http://arxiv.org/abs/1009.0947}{{\tt arXiv:1009.0947}}].

\bibitem{Gorbahn:2006bm}
M.~Gorbahn and U.~Haisch, {\it {Charm quark contribution to $K_L \to \mu^+
  \mu^-$ at next-to-next-to-leading order}},  {\em Phys.~Rev.~Lett.} {\bf 97}
  (2006) 122002, [\href{http://arxiv.org/abs/hep-ph/0605203}{{\tt
  hep-ph/0605203}}].

\bibitem{Buchalla:1995vs}
G.~Buchalla, A.~J. Buras, and M.~E. Lautenbacher, {\it {Weak decays beyond
  leading logarithms}},  {\em Rev.~Mod.~Phys.} {\bf 68} (1996) 1125--1144,
  [\href{http://arxiv.org/abs/hep-ph/9512380}{{\tt hep-ph/9512380}}].

\bibitem{Buras:2011we}
A.~J. Buras, {\it {Climbing NLO and NNLO Summits of Weak Decays}},
  \href{http://arxiv.org/abs/1102.5650}{{\tt arXiv:1102.5650}}.

\bibitem{Christ:2016eae}
{\bf RBC, UKQCD} Collaboration, N.~H. Christ, X.~Feng, A.~Portelli, and C.~T.
  Sachrajda, {\it {Prospects for a lattice computation of rare kaon decay
  amplitudes II $K\to\pi\nu\bar{\nu}$ decays}},  {\em Phys. Rev.} {\bf D93}
  (2016), no.~11 114517, [\href{http://arxiv.org/abs/1605.04442}{{\tt
  arXiv:1605.04442}}].

\bibitem{Christ:2019dxu}
{\bf RBC, UKQCD} Collaboration, N.~H. Christ, X.~Feng, A.~Portelli, and C.~T.
  Sachrajda, {\it {Lattice QCD study of the rare kaon decay
  $K^+\to\pi^+\nu\bar{\nu}$ at a near-physical pion mass}},  {\em Phys. Rev. D}
  {\bf 100} (2019), no.~11 114506, [\href{http://arxiv.org/abs/1910.10644}{{\tt
  arXiv:1910.10644}}].

\bibitem{Li:2019fhz}
T.~Li, X.-D. Ma, and M.~A. Schmidt, {\it {Implication of $K\to \pi \nu
  \bar{\nu}$ for generic neutrino interactions in effective field theories}},
  {\em Phys. Rev. D} {\bf 101} (2020), no.~5 055019,
  [\href{http://arxiv.org/abs/1912.10433}{{\tt arXiv:1912.10433}}].

\bibitem{Deppisch:2020oyx}
F.~F. Deppisch, K.~Fridell, and J.~Harz, {\it {Probing lepton number violating
  interactions in rare kaon decays}},  {\em JHEP} {\bf 12} (2020) 186,
  [\href{http://arxiv.org/abs/2009.04494}{{\tt arXiv:2009.04494}}].

\bibitem{Buras:2004uu}
A.~J. Buras, F.~Schwab, and S.~Uhlig, {\it {Waiting for precise measurements of
  $K^{+} \to \pi^{+} \nu \bar{\nu}$ and $K_{L} \to \pi^0 \nu \bar{\nu}$}},
  {\em Rev.~Mod.~Phys.} {\bf 80} (2008) 965--1007,
  [\href{http://arxiv.org/abs/hep-ph/0405132}{{\tt hep-ph/0405132}}].

\bibitem{Komatsubara:2012pn}
T.~Komatsubara, {\it {Experiments with K-Meson Decays}},  {\em
  Prog.~Part.~Nucl.~Phys.} {\bf 67} (2012) 995--1018,
  [\href{http://arxiv.org/abs/1203.6437}{{\tt arXiv:1203.6437}}].

\bibitem{Buras:2013ooa}
A.~J. Buras and J.~Girrbach, {\it {Towards the Identification of New Physics
  through Quark Flavour Violating Processes}},  {\em Rept.~Prog.~Phys.} {\bf
  77} (2014) 086201, [\href{http://arxiv.org/abs/1306.3775}{{\tt
  arXiv:1306.3775}}].

\bibitem{Blanke:2013goa}
M.~Blanke, {\it {New Physics Signatures in Kaon Decays}},  {\em PoS} {\bf
  KAON13} (2013) 010, [\href{http://arxiv.org/abs/1305.5671}{{\tt
  arXiv:1305.5671}}].

\bibitem{Smith:2014mla}
C.~Smith, {\it {Rare K decays: Challenges and Perspectives}},
  \href{http://arxiv.org/abs/1409.6162}{{\tt arXiv:1409.6162}}.

\bibitem{Buras:2014zga}
A.~J. Buras, D.~Buttazzo, J.~Girrbach-Noe, and R.~Knegjens, {\it {Can we reach
  the Zeptouniverse with rare $K$ and $B_{s,d}$ decays?}},  {\em JHEP} {\bf
  1411} (2014) 121, [\href{http://arxiv.org/abs/1408.0728}{{\tt
  arXiv:1408.0728}}].

\bibitem{Buras:2022hws}
A.~J. Buras and J.~Harz, {\it {Majorana Footprints}},
  \href{http://arxiv.org/abs/2204.XXXXX}{{\tt arXiv:2204.XXXXX}}.

\bibitem{Goudzovski:2022vbt}
E.~Goudzovski et~al., {\it {New Physics Searches at Kaon and Hyperon
  Factories}},  \href{http://arxiv.org/abs/2201.07805}{{\tt arXiv:2201.07805}}.

\bibitem{Brod:2021hsj}
J.~Brod, M.~Gorbahn, and E.~Stamou, {\it {Updated Standard Model Prediction for
  $K \to \pi \nu \bar{\nu}$ and $\epsilon_K$}},  in {\em {19th International
  Conference on B-Physics at Frontier Machines}}, 5, 2021.
\newblock \href{http://arxiv.org/abs/2105.02868}{{\tt arXiv:2105.02868}}.

\bibitem{Buras:2012jb}
A.~J. Buras, F.~De~Fazio, and J.~Girrbach, {\it {The Anatomy of Z' and Z with
  Flavour Changing Neutral Currents in the Flavour Precision Era}},  {\em JHEP}
  {\bf 1302} (2013) 116, [\href{http://arxiv.org/abs/1211.1896}{{\tt
  arXiv:1211.1896}}].

\bibitem{Buras:2015qea}
A.~J. Buras, D.~Buttazzo, J.~Girrbach-Noe, and R.~Knegjens, {\it {$ {K}^{+}\to
  {\pi}^{+}\nu \overline{\nu} $ and $ {K}_L\to {\pi}^0\nu \overline{\nu} $ in
  the Standard Model: status and perspectives}},  {\em JHEP} {\bf 11} (2015)
  033, [\href{http://arxiv.org/abs/1503.02693}{{\tt arXiv:1503.02693}}].

\bibitem{Bobeth:2016llm}
C.~Bobeth, A.~J. Buras, A.~Celis, and M.~Jung, {\it {Patterns of Flavour
  Violation in Models with Vector-Like Quarks}},  {\em JHEP} {\bf 04} (2017)
  079, [\href{http://arxiv.org/abs/1609.04783}{{\tt arXiv:1609.04783}}].

\bibitem{NA62:2020upd}
{\bf NA62, KLEVER} Collaboration, {\it {Rare decays at the CERN high-intensity
  kaon beam facility}},  \href{http://arxiv.org/abs/2009.10941}{{\tt
  arXiv:2009.10941}}.

\bibitem{CortinaGil:2020vlo}
{\bf NA62} Collaboration, E.~Cortina~Gil et~al., {\it {An investigation of the
  very rare $K^+\rightarrow\pi^+\nu\bar{\nu}$ decay}},
  \href{http://arxiv.org/abs/2007.08218}{{\tt arXiv:2007.08218}}.

\bibitem{NA62ichep}
R.~Marchevski, ``{ New result on the search for the $K^+ \to \pi^+ \nu \bar
  \nu$ decay at the NA62 experiment at CERN}.'' ICHEP, Prague, 28 July-6
  August, 2020.

\bibitem{Zyla:2020zbs}
{\bf Particle Data Group} Collaboration, P.~A. Zyla et~al., {\it {Review of
  Particle Physics}},  {\em PTEP} {\bf 2020} (2020), no.~8 083C01.

\bibitem{Aoki:2021kgd}
Y.~Aoki et~al., {\it {FLAG Review 2021}},
  \href{http://arxiv.org/abs/2111.09849}{{\tt arXiv:2111.09849}}.

\bibitem{Dowdall:2019bea}
R.~J. Dowdall, C.~T.~H. Davies, R.~R. Horgan, G.~P. Lepage, C.~J. Monahan,
  J.~Shigemitsu, and M.~Wingate, {\it {Neutral $B$-meson mixing from full
  lattice QCD at the physical point}},  {\em Phys. Rev. D} {\bf 100} (2019),
  no.~9 094508, [\href{http://arxiv.org/abs/1907.01025}{{\tt
  arXiv:1907.01025}}].

\bibitem{Brod:2019rzc}
J.~Brod, M.~Gorbahn, and E.~Stamou, {\it {Standard-Model Prediction of
  $\epsilon_K$ with Manifest Quark-Mixing Unitarity}},  {\em Phys. Rev. Lett.}
  {\bf 125} (2020), no.~17 171803, [\href{http://arxiv.org/abs/1911.06822}{{\tt
  arXiv:1911.06822}}].

\bibitem{Buras:2010pza}
A.~J. Buras, D.~Guadagnoli, and G.~Isidori, {\it {On $\epsilon_K$ beyond lowest
  order in the Operator Product Expansion}},  {\em Phys.~Lett.} {\bf B688}
  (2010) 309--313, [\href{http://arxiv.org/abs/1002.3612}{{\tt
  arXiv:1002.3612}}].

\bibitem{Buras:1990fn}
A.~J. Buras, M.~Jamin, and P.~H. Weisz, {\it {Leading and next-to-leading QCD
  corrections to $\varepsilon$ parameter and $B^0-\bar{B}^0$ mixing in the
  presence of a heavy top quark}},  {\em Nucl.~Phys.} {\bf B347} (1990)
  491--536.

\bibitem{Urban:1997gw}
J.~Urban, F.~Krauss, U.~Jentschura, and G.~Soff, {\it {Next-to-leading order
  QCD corrections for the $B^0 - \bar B^0$ mixing with an extended Higgs
  sector}},  {\em Nucl.~Phys.} {\bf B523} (1998) 40--58,
  [\href{http://arxiv.org/abs/hep-ph/9710245}{{\tt hep-ph/9710245}}].

\bibitem{Buchalla:1996fp}
G.~Buchalla and A.~J. Buras, {\it {$K \to\pi\nu\bar\nu$ and high precision
  determinations of the CKM matrix}},  {\em Phys.~Rev.} {\bf D54} (1996)
  6782--6789, [\href{http://arxiv.org/abs/hep-ph/9607447}{{\tt
  hep-ph/9607447}}].

\bibitem{Shiomi:2014sfa}
{\bf KOTO} Collaboration, K.~Shiomi, {\it {$K^{0}_{L}\rightarrow \pi^{0} \nu
  \bar{\nu}$ at KOTO}},  in {\em {8th International Workshop on the CKM
  Unitarity Triangle (CKM 2014) Vienna, Austria, September 8-12, 2014}}, 2014.
\newblock \href{http://arxiv.org/abs/1411.4250}{{\tt arXiv:1411.4250}}.

\bibitem{Grossman:1997sk}
Y.~Grossman and Y.~Nir, {\it {$K_L\to\pi^0\nu\bar\nu$ beyond the standard
  model}},  {\em Phys.~Lett.} {\bf B398} (1997) 163--168,
  [\href{http://arxiv.org/abs/hep-ph/9701313}{{\tt hep-ph/9701313}}].

\bibitem{Kitahara:2019lws}
T.~Kitahara, T.~Okui, G.~Perez, Y.~Soreq, and K.~Tobioka, {\it {New physics
  implications of recent search for $K_L \to \pi^0 \nu\bar{\nu}$ at KOTO}},
  {\em Phys. Rev. Lett.} {\bf 124} (2020), no.~7 071801,
  [\href{http://arxiv.org/abs/1909.11111}{{\tt arXiv:1909.11111}}].

\bibitem{He:2020jly}
X.-G. He, X.-D. Ma, J.~Tandean, and G.~Valencia, {\it {Evading the Grossman-Nir
  bound with $\Delta I=3/2$ new physics}},  {\em JHEP} {\bf 08} (2020), no.~08
  034, [\href{http://arxiv.org/abs/2005.02942}{{\tt arXiv:2005.02942}}].

\bibitem{He:2020jzn}
X.-G. He, X.-D. Ma, J.~Tandean, and G.~Valencia, {\it {Breaking the
  Grossman-Nir Bound in Kaon Decays}},  {\em JHEP} {\bf 04} (2020) 057,
  [\href{http://arxiv.org/abs/2002.05467}{{\tt arXiv:2002.05467}}].

\bibitem{Fuyuto:2014cya}
K.~Fuyuto, W.-S. Hou, and M.~Kohda, {\it {Loophole in $K \to \pi\nu\bar{\nu}$
  Search and New Weak Leptonic Forces}},  {\em Phys. Rev. Lett.} {\bf 114}
  (2015) 171802, [\href{http://arxiv.org/abs/1412.4397}{{\tt
  arXiv:1412.4397}}].

\bibitem{Buras:2015yca}
A.~J. Buras, D.~Buttazzo, and R.~Knegjens, {\it {$K\to\pi\nu\bar\nu$ and
  $\epsilon'/\epsilon$ in Simplified New Physics Models}},  {\em JHEP} {\bf 11}
  (2015) 166, [\href{http://arxiv.org/abs/1507.08672}{{\tt arXiv:1507.08672}}].

\bibitem{Buras:2001af}
A.~J. Buras and R.~Fleischer, {\it {Bounds on the unitarity triangle,
  $\sin2\beta$ and $K \to\pi \nu\bar\nu$ decays in models with minimal flavor
  violation}},  {\em Phys.~Rev.} {\bf D64} (2001) 115010,
  [\href{http://arxiv.org/abs/hep-ph/0104238}{{\tt hep-ph/0104238}}].

\bibitem{Blanke:2009pq}
M.~Blanke, {\it {Insights from the Interplay of $K\rightarrow \pi
  \nu\overline{\nu}$ and $\epsilon_K$ on the New Physics Flavour Structure}},
  {\em Acta Phys.Polon.} {\bf B41} (2010) 127,
  [\href{http://arxiv.org/abs/0904.2528}{{\tt arXiv:0904.2528}}].

\bibitem{Barbieri:2014tja}
R.~Barbieri, D.~Buttazzo, F.~Sala, and D.~M. Straub, {\it {Flavour physics and
  flavour symmetries after the first LHC phase}},  {\em JHEP} {\bf 1405} (2014)
  105, [\href{http://arxiv.org/abs/1402.6677}{{\tt arXiv:1402.6677}}].

\bibitem{Blanke:2009am}
M.~Blanke, A.~J. Buras, B.~Duling, S.~Recksiegel, and C.~Tarantino, {\it {FCNC
  Processes in the Littlest Higgs Model with T-Parity: a 2009 Look}},  {\em
  Acta Phys.Polon.} {\bf B41} (2010) 657--683,
  [\href{http://arxiv.org/abs/0906.5454}{{\tt arXiv:0906.5454}}].

\bibitem{Blanke:2008yr}
M.~Blanke, A.~J. Buras, B.~Duling, K.~Gemmler, and S.~Gori, {\it {Rare K and B
  Decays in a Warped Extra Dimension with Custodial Protection}},  {\em JHEP}
  {\bf 03} (2009) 108, [\href{http://arxiv.org/abs/0812.3803}{{\tt
  arXiv:0812.3803}}].

\bibitem{Bobeth:2013tba}
C.~Bobeth, M.~Gorbahn, and E.~Stamou, {\it {Electroweak Corrections to $B_{s,d}
  \to \ell^+ \ell^-$}},  {\em Phys.~Rev.} {\bf D89} (2014) 034023,
  [\href{http://arxiv.org/abs/1311.1348}{{\tt arXiv:1311.1348}}].

\bibitem{Ecker:1991ru}
G.~Ecker and A.~Pich, {\it {The Longitudinal muon polarization in $K_L \to
  \mu^+ \mu^-$}},  {\em Nucl. Phys.} {\bf B366} (1991) 189--205.

\bibitem{Aaij:2017tia}
{\bf LHCb} Collaboration, R.~Aaij et~al., {\it {Improved limit on the branching
  fraction of the rare decay ${{K} ^0_{\mathrm { \scriptscriptstyle S}}}
  \rightarrow \mu ^+\mu ^-$}},  {\em Eur. Phys. J.} {\bf C77} (2017), no.~10
  678, [\href{http://arxiv.org/abs/1706.00758}{{\tt arXiv:1706.00758}}].

\bibitem{Dery:2021vql}
A.~Dery and M.~Ghosh, {\it {$K\to\mu^+\mu^-$ beyond the standard model}},
  \href{http://arxiv.org/abs/2112.05801}{{\tt arXiv:2112.05801}}.

\bibitem{Bobeth:2017ecx}
C.~Bobeth and A.~J. Buras, {\it {Leptoquarks meet $\varepsilon'/\varepsilon$
  and rare Kaon processes}},  {\em JHEP} {\bf 02} (2018) 101,
  [\href{http://arxiv.org/abs/1712.01295}{{\tt arXiv:1712.01295}}].

\bibitem{D'Ambrosio:1998yj}
G.~D'Ambrosio, G.~Ecker, G.~Isidori, and J.~Portoles, {\it {The decays $K \to
  \pi \ell^+ \ell^-$ beyond leading order in the chiral expansion}},  {\em
  JHEP} {\bf 08} (1998) 004, [\href{http://arxiv.org/abs/hep-ph/9808289}{{\tt
  hep-ph/9808289}}].

\bibitem{Isidori:2004rb}
G.~Isidori, C.~Smith, and R.~Unterdorfer, {\it {The rare decay $K_L \to \pi^0
  \mu^+\mu^-$ within the SM}},  {\em Eur. Phys. J.} {\bf C36} (2004) 57--66,
  [\href{http://arxiv.org/abs/hep-ph/0404127}{{\tt hep-ph/0404127}}].

\bibitem{Friot:2004yr}
S.~Friot, D.~Greynat, and E.~De~Rafael, {\it {Rare kaon decays revisited}},
  {\em Phys.~Lett.} {\bf B595} (2004) 301--308,
  [\href{http://arxiv.org/abs/hep-ph/0404136}{{\tt hep-ph/0404136}}].

\bibitem{Alonso:2014csa}
R.~Alonso, B.~Grinstein, and J.~Martin~Camalich, {\it {$SU(2)\times U(1)$ gauge
  invariance and the shape of new physics in rare $B$ decays}},  {\em Phys.
  Rev. Lett.} {\bf 113} (2014) 241802,
  [\href{http://arxiv.org/abs/1407.7044}{{\tt arXiv:1407.7044}}].

\bibitem{Buchalla:1990qz}
G.~Buchalla, A.~J. Buras, and M.~K. Harlander, {\it {Penguin box expansion:
  Flavor changing neutral current processes and a heavy top quark}},  {\em
  Nucl.~Phys.} {\bf B349} (1991) 1--47.

\bibitem{AlaviHarati:2003mr}
{\bf KTeV} Collaboration, A.~Alavi-Harati et~al., {\it {Search for the rare
  decay $K_L\to\pi^0 e^+ e^-$}},  {\em Phys. Rev. Lett.} {\bf 93} (2004)
  021805, [\href{http://arxiv.org/abs/hep-ex/0309072}{{\tt hep-ex/0309072}}].

\bibitem{AlaviHarati:2000hs}
{\bf KTEV} Collaboration, A.~Alavi-Harati et~al., {\it {Search for the Decay
  $K_L \to \pi^0 \mu^+ \mu^-$}},  {\em Phys. Rev. Lett.} {\bf 84} (2000)
  5279--5282, [\href{http://arxiv.org/abs/hep-ex/0001006}{{\tt
  hep-ex/0001006}}].

\bibitem{Prades:2007ud}
J.~Prades, {\it {ChPT Progress on Non-Leptonic and Radiative Kaon Decays}},
  {\em PoS} {\bf KAON} (2008) 022, [\href{http://arxiv.org/abs/0707.1789}{{\tt
  arXiv:0707.1789}}].

\bibitem{Bruno:1992za}
C.~Bruno and J.~Prades, {\it {Rare Kaon Decays in the $1/N_c$-Expansion}},
  {\em Z. Phys.} {\bf C57} (1993) 585--594,
  [\href{http://arxiv.org/abs/hep-ph/9209231}{{\tt hep-ph/9209231}}].

\bibitem{Buras:2013rqa}
A.~J. Buras, F.~De~Fazio, J.~Girrbach, R.~Knegjens, and M.~Nagai, {\it {The
  Anatomy of Neutral Scalars with FCNCs in the Flavour Precision Era}},  {\em
  JHEP} {\bf 1306} (2013) 111, [\href{http://arxiv.org/abs/1303.3723}{{\tt
  arXiv:1303.3723}}].

\bibitem{Buras:2003td}
A.~J. Buras, {\it {Relations between $\Delta M_{s,d}$ and $B_{s,d} \to \mu^+
  \mu^-$ in models with minimal flavour violation}},  {\em Phys.~Lett.} {\bf
  B566} (2003) 115--119, [\href{http://arxiv.org/abs/hep-ph/0303060}{{\tt
  hep-ph/0303060}}].

\bibitem{Bobeth:2021cxm}
C.~Bobeth and A.~J. Buras, {\it {Searching for New Physics with
  $\overline{\mathcal{B}}(B_{s,d}\to\mu\bar\mu)/\Delta M_{s,d}$}},  {\em Acta
  Phys. Polon. B} {\bf 52} (2021) 1189,
  [\href{http://arxiv.org/abs/2104.09521}{{\tt arXiv:2104.09521}}].

\bibitem{Brod:2021qvc}
J.~Brod, S.~Kvedarait\.{e}, and Z.~Polonsky, {\it {Two-loop electroweak
  corrections to the Top-Quark Contribution to $\epsilon_K$}},  {\em JHEP} {\bf
  12} (2021) 198, [\href{http://arxiv.org/abs/2108.00017}{{\tt
  arXiv:2108.00017}}].

\bibitem{Buras:2015jaq}
A.~J. Buras, {\it {New physics patterns in $\varepsilon^\prime/\varepsilon$ and
  $\varepsilon_K$ with implications for rare kaon decays and $\Delta M_K$}},
  {\em JHEP} {\bf 04} (2016) 071, [\href{http://arxiv.org/abs/1601.00005}{{\tt
  arXiv:1601.00005}}].

\bibitem{Aebischer:2020mkv}
J.~Aebischer, A.~J. Buras, and J.~Kumar, {\it {Another SMEFT story: $Z^\prime$
  facing new results on $\epsilon'/\epsilon$, $\Delta M_{K}$ and $K \to \pi \nu
  \overline{\nu} $}},  {\em JHEP} {\bf 12} (2020) 097,
  [\href{http://arxiv.org/abs/2006.01138}{{\tt arXiv:2006.01138}}].

\bibitem{Altmannshofer:2009ma}
W.~Altmannshofer, A.~J. Buras, D.~M. Straub, and M.~Wick, {\it {New strategies
  for New Physics search in $B \to K^{*} \nu \bar{\nu}$, $B \to K \nu
  \bar{\nu}$ and $B \to X_{s} \nu \bar{\nu}$ decays}},  {\em JHEP} {\bf 04}
  (2009) 022, [\href{http://arxiv.org/abs/0902.0160}{{\tt arXiv:0902.0160}}].

\bibitem{Buras:2014fpa}
A.~J. Buras, J.~Girrbach-Noe, C.~Niehoff, and D.~M. Straub, {\it {$B\to
  K^{(*)}\nu\bar\nu$ decays in the Standard Model and beyond}},  {\em JHEP}
  {\bf 1502} (2015) 184, [\href{http://arxiv.org/abs/1409.4557}{{\tt
  arXiv:1409.4557}}].

\bibitem{Bause:2021cna}
R.~Bause, H.~Gisbert, M.~Golz, and G.~Hiller, {\it {Interplay of dineutrino
  modes with semileptonic rare B-decays}},  {\em JHEP} {\bf 12} (2021) 061,
  [\href{http://arxiv.org/abs/2109.01675}{{\tt arXiv:2109.01675}}].

\bibitem{Buras:2012dp}
A.~J. Buras, F.~De~Fazio, J.~Girrbach, and M.~V. Carlucci, {\it {The Anatomy of
  Quark Flavour Observables in 331 Models in the Flavour Precision Era}},  {\em
  JHEP} {\bf 1302} (2013) 023, [\href{http://arxiv.org/abs/1211.1237}{{\tt
  arXiv:1211.1237}}].

\bibitem{Aebischer:2019blw}
J.~Aebischer, A.~J. Buras, M.~Cerd{\'a}-Sevilla, and F.~De~Fazio, {\it
  {Quark-lepton connections in Z' mediated FCNC processes: gauge anomaly
  cancellations at work}},  {\em JHEP} {\bf 02} (2020) 183,
  [\href{http://arxiv.org/abs/1912.09308}{{\tt arXiv:1912.09308}}].

\bibitem{Bobeth:2017xry}
C.~Bobeth, A.~J. Buras, A.~Celis, and M.~Jung, {\it {Yukawa enhancement of
  $Z$-mediated new physics in $\Delta S = 2$ and $\Delta B = 2$ processes}},
  {\em JHEP} {\bf 07} (2017) 124, [\href{http://arxiv.org/abs/1703.04753}{{\tt
  arXiv:1703.04753}}].

\bibitem{Endo:2018gdn}
M.~Endo, T.~Kitahara, and D.~Ueda, {\it {SMEFT top-quark effects on $\Delta
  F=2$ observables}},  {\em JHEP} {\bf 07} (2019) 182,
  [\href{http://arxiv.org/abs/1811.04961}{{\tt arXiv:1811.04961}}].

\bibitem{Fajfer:2018bfj}
S.~Fajfer, N.~Košnik, and L.~Vale~Silva, {\it {Footprints of leptoquarks: from
  $ R_{K^{(*)}} $ to $ K \rightarrow \pi \nu \bar{\nu }$}},  {\em Eur. Phys.
  J.} {\bf C78} (2018), no.~4 275, [\href{http://arxiv.org/abs/1802.00786}{{\tt
  arXiv:1802.00786}}].

\bibitem{Marzocca:2021miv}
D.~Marzocca, S.~Trifinopoulos, and E.~Venturini, {\it {From B-meson anomalies
  to Kaon physics with scalar leptoquarks}},
  \href{http://arxiv.org/abs/2106.15630}{{\tt arXiv:2106.15630}}.

\bibitem{Bordone:2017lsy}
M.~Bordone, D.~Buttazzo, G.~Isidori, and J.~Monnard, {\it {Probing Lepton
  Flavour Universality with $K \to \pi \nu \bar\nu$ decays}},  {\em Eur. Phys.
  J.} {\bf C77} (2017), no.~9 618, [\href{http://arxiv.org/abs/1705.10729}{{\tt
  arXiv:1705.10729}}].

\bibitem{Crivellin:2016vjc}
A.~Crivellin, G.~D'Ambrosio, M.~Hoferichter, and L.~C. Tunstall, {\it
  {Violation of lepton flavor and lepton flavor universality in rare kaon
  decays}},  {\em Phys. Rev.} {\bf D93} (2016), no.~7 074038,
  [\href{http://arxiv.org/abs/1601.00970}{{\tt arXiv:1601.00970}}].

\bibitem{Endo:2016tnu}
M.~Endo, T.~Kitahara, S.~Mishima, and K.~Yamamoto, {\it {Revisiting Kaon
  Physics in General $Z$ Scenario}},  {\em Phys. Lett.} {\bf B771} (2017)
  37--44, [\href{http://arxiv.org/abs/1612.08839}{{\tt arXiv:1612.08839}}].

\bibitem{Blanke:2006eb}
M.~Blanke et~al., {\it {Rare and CP-violating $K$ and $B$ decays in the
  Littlest Higgs model with T-parity}},  {\em JHEP} {\bf 01} (2007) 066,
  [\href{http://arxiv.org/abs/hep-ph/0610298}{{\tt hep-ph/0610298}}].

\bibitem{Buras:2006wk}
A.~J. Buras, A.~Poschenrieder, S.~Uhlig, and W.~A. Bardeen, {\it {Rare $K$ and
  $B$ decays in the Littlest Higgs model without T-parity}},  {\em JHEP} {\bf
  11} (2006) 062, [\href{http://arxiv.org/abs/hep-ph/0607189}{{\tt
  hep-ph/0607189}}].

\bibitem{Blanke:2015wba}
M.~Blanke, A.~J. Buras, and S.~Recksiegel, {\it {Quark flavour observables in
  the Littlest Higgs model with T-parity after LHC Run 1}},  {\em Eur. Phys.
  J.} {\bf C76} (2016), no.~4 182, [\href{http://arxiv.org/abs/1507.06316}{{\tt
  arXiv:1507.06316}}].

\bibitem{Buras:2004qb}
A.~J. Buras, T.~Ewerth, S.~Jager, and J.~Rosiek, {\it {$K^+ \to \pi^+ \nu
  \bar\nu$ and $K_L \to \pi^0 \nu \bar\nu$ decays in the general MSSM}},  {\em
  Nucl.~Phys.} {\bf B714} (2005) 103--136,
  [\href{http://arxiv.org/abs/hep-ph/0408142}{{\tt hep-ph/0408142}}].

\bibitem{Isidori:2006qy}
G.~Isidori, F.~Mescia, P.~Paradisi, C.~Smith, and S.~Trine, {\it {Exploring the
  flavour structure of the MSSM with rare K decays}},  {\em JHEP} {\bf 08}
  (2006) 064, [\href{http://arxiv.org/abs/hep-ph/0604074}{{\tt
  hep-ph/0604074}}].

\bibitem{Blazek:2014qda}
T.~Blazek and P.~Matak, {\it {Left--left squark mixing, $K^+\rightarrow
  \pi^+\nu\bar{\nu}$ and minimal supersymmetry with large tan$\beta$}},  {\em
  Int. J. Mod. Phys. A} {\bf 29} (2014), no.~27 1450162,
  [\href{http://arxiv.org/abs/1410.0055}{{\tt arXiv:1410.0055}}].

\bibitem{Altmannshofer:2009ne}
W.~Altmannshofer, A.~J. Buras, S.~Gori, P.~Paradisi, and D.~M. Straub, {\it
  {Anatomy and Phenomenology of FCNC and CPV Effects in SUSY Theories}},  {\em
  Nucl.~Phys.} {\bf B830} (2010) 17--94,
  [\href{http://arxiv.org/abs/0909.1333}{{\tt arXiv:0909.1333}}].

\bibitem{Tanimoto:2016yfy}
M.~Tanimoto and K.~Yamamoto, {\it {Probing SUSY with 10 TeV stop mass in rare
  decays and CP violation of kaon}},  {\em PTEP} {\bf 2016} (2016), no.~12
  123B02, [\href{http://arxiv.org/abs/1603.07960}{{\tt arXiv:1603.07960}}].

\bibitem{Kitahara:2016otd}
T.~Kitahara, U.~Nierste, and P.~Tremper, {\it {Supersymmetric Explanation of CP
  Violation in $K\to \pi\pi$ Decays}},  {\em Phys. Rev. Lett.} {\bf 117}
  (2016), no.~9 091802, [\href{http://arxiv.org/abs/1604.07400}{{\tt
  arXiv:1604.07400}}].

\bibitem{Endo:2016aws}
M.~Endo, S.~Mishima, D.~Ueda, and K.~Yamamoto, {\it {Chargino contributions in
  light of recent $\epsilon'/\epsilon$}},  {\em Phys. Lett.} {\bf B762} (2016)
  493--497, [\href{http://arxiv.org/abs/1608.01444}{{\tt arXiv:1608.01444}}].

\bibitem{Crivellin:2017gks}
A.~Crivellin, G.~D'Ambrosio, T.~Kitahara, and U.~Nierste, {\it {$K\to \pi
  \nu\overline{\nu}$ in the MSSM in light of the
  $\epsilon^{\prime}_K/\epsilon_K$ anomaly}},  {\em Phys. Rev.} {\bf D96}
  (2017), no.~1 015023, [\href{http://arxiv.org/abs/1703.05786}{{\tt
  arXiv:1703.05786}}].

\bibitem{Endo:2017ums}
M.~Endo, T.~Goto, T.~Kitahara, S.~Mishima, D.~Ueda, and K.~Yamamoto, {\it
  {Gluino-mediated electroweak penguin with flavor-violating trilinear
  couplings}},  {\em JHEP} {\bf 04} (2018) 019,
  [\href{http://arxiv.org/abs/1712.04959}{{\tt arXiv:1712.04959}}].

\bibitem{Chen:2018ytc}
C.-H. Chen and T.~Nomura, {\it {$Re(\epsilon'_K/\epsilon_K$) and $K \to \pi \nu
  \bar\nu$ in a two-Higgs doublet model}},  {\em JHEP} {\bf 08} (2018) 145,
  [\href{http://arxiv.org/abs/1804.06017}{{\tt arXiv:1804.06017}}].

\bibitem{Chen:2018vog}
C.-H. Chen and T.~Nomura, {\it {$\epsilon'/\epsilon$ from charged-Higgs-induced
  gluonic dipole operators}},  {\em Phys. Lett.} {\bf B787} (2018) 182--187,
  [\href{http://arxiv.org/abs/1805.07522}{{\tt arXiv:1805.07522}}].

\bibitem{Buras:2002ej}
A.~J. Buras, M.~Spranger, and A.~Weiler, {\it {The impact of universal extra
  dimensions on the unitarity triangle and rare $K$ and $B$ decays}},  {\em
  Nucl.~Phys.} {\bf B660} (2003) 225--268,
  [\href{http://arxiv.org/abs/hep-ph/0212143}{{\tt hep-ph/0212143}}].

\bibitem{Buras:2003mk}
A.~J. Buras, A.~Poschenrieder, M.~Spranger, and A.~Weiler, {\it {The impact of
  universal extra dimensions on $B \to X_s \gamma$, $B \to X_s$gluon, $B \to
  X_s \mu^+ \mu^-$, $K_L \to\pi^0 e^+ e^-$, and $\varepsilon'/\varepsilon$}},
  {\em Nucl.~Phys.} {\bf B678} (2004) 455--490,
  [\href{http://arxiv.org/abs/hep-ph/0306158}{{\tt hep-ph/0306158}}].

\bibitem{Blanke:2008zb}
M.~Blanke, A.~J. Buras, B.~Duling, S.~Gori, and A.~Weiler, {\it {$\Delta F=2$
  Observables and Fine-Tuning in a Warped Extra Dimension with Custodial
  Protection}},  {\em JHEP} {\bf 03} (2009) 001,
  [\href{http://arxiv.org/abs/0809.1073}{{\tt arXiv:0809.1073}}].

\bibitem{Albrecht:2009xr}
M.~E. Albrecht, M.~Blanke, A.~J. Buras, B.~Duling, and K.~Gemmler, {\it
  {Electroweak and Flavour Structure of a Warped Extra Dimension with Custodial
  Protection}},  {\em JHEP} {\bf 09} (2009) 064,
  [\href{http://arxiv.org/abs/0903.2415}{{\tt arXiv:0903.2415}}].

\bibitem{Casagrande:2008hr}
S.~Casagrande, F.~Goertz, U.~Haisch, M.~Neubert, and T.~Pfoh, {\it {Flavor
  Physics in the Randall-Sundrum Model: I. Theoretical Setup and Electroweak
  Precision Tests}},  {\em JHEP} {\bf 10} (2008) 094,
  [\href{http://arxiv.org/abs/0807.4937}{{\tt arXiv:0807.4937}}].

\bibitem{Bauer:2009cf}
M.~Bauer, S.~Casagrande, U.~Haisch, and M.~Neubert, {\it {Flavor Physics in the
  Randall-Sundrum Model: II. Tree-Level Weak-Interaction Processes}},  {\em
  JHEP} {\bf 1009} (2010) 017, [\href{http://arxiv.org/abs/0912.1625}{{\tt
  arXiv:0912.1625}}].

\bibitem{Aebischer:2018csl}
J.~Aebischer, C.~Bobeth, A.~J. Buras, and D.~M. Straub, {\it {Anatomy of
  $\varepsilon '/\varepsilon $ beyond the standard model}},  {\em Eur. Phys.
  J.} {\bf C79} (2019), no.~3 219, [\href{http://arxiv.org/abs/1808.00466}{{\tt
  arXiv:1808.00466}}].

\bibitem{Matsuzaki:2018jui}
S.~Matsuzaki, K.~Nishiwaki, and K.~Yamamoto, {\it {Simultaneous interpretation
  of $K$ and $B$ anomalies in terms of chiral-flavorful vectors}},  {\em JHEP}
  {\bf 11} (2018) 164, [\href{http://arxiv.org/abs/1806.02312}{{\tt
  arXiv:1806.02312}}].

\bibitem{Chen:2018dfc}
C.-H. Chen and T.~Nomura, {\it {$\epsilon_K$ and $\epsilon'/\epsilon$ in a
  diquark model}},  {\em JHEP} {\bf 03} (2019) 009,
  [\href{http://arxiv.org/abs/1808.04097}{{\tt arXiv:1808.04097}}].

\bibitem{Chen:2018stt}
C.-H. Chen and T.~Nomura, {\it {Left-handed color-sextet diquark in the Kaon
  system}},  {\em Phys. Rev.} {\bf D99} (2019), no.~11 115006,
  [\href{http://arxiv.org/abs/1811.02315}{{\tt arXiv:1811.02315}}].

\bibitem{Matsuzaki:2019clv}
S.~Matsuzaki, K.~Nishiwaki, and K.~Yamamoto, {\it {Simultaneous explanation of
  $K$ and $B$ anomalies in vectorlike compositeness}},  in {\em {18th Hellenic
  School and Workshops on Elementary Particle Physics and Gravity (CORFU2018)
  Corfu, Corfu, Greece, August 31-September 28, 2018}}, 2019.
\newblock \href{http://arxiv.org/abs/1903.10823}{{\tt arXiv:1903.10823}}.

\bibitem{Grzadkowski:2010es}
B.~Grzadkowski, M.~Iskrzynski, M.~Misiak, and J.~Rosiek, {\it {Dimension-Six
  Terms in the Standard Model Lagrangian}},  {\em JHEP} {\bf 1010} (2010) 085,
  [\href{http://arxiv.org/abs/1008.4884}{{\tt arXiv:1008.4884}}].

\bibitem{Jenkins:2013zja}
E.~E. Jenkins, A.~V. Manohar, and M.~Trott, {\it {Renormalization Group
  Evolution of the Standard Model Dimension Six Operators I: Formalism and
  lambda Dependence}},  {\em JHEP} {\bf 10} (2013) 087,
  [\href{http://arxiv.org/abs/1308.2627}{{\tt arXiv:1308.2627}}].

\bibitem{Jenkins:2013wua}
E.~E. Jenkins, A.~V. Manohar, and M.~Trott, {\it {Renormalization Group
  Evolution of the Standard Model Dimension Six Operators II: Yukawa
  Dependence}},  {\em JHEP} {\bf 01} (2014) 035,
  [\href{http://arxiv.org/abs/1310.4838}{{\tt arXiv:1310.4838}}].

\bibitem{Alonso:2013hga}
R.~Alonso, E.~E. Jenkins, A.~V. Manohar, and M.~Trott, {\it {Renormalization
  Group Evolution of the Standard Model Dimension Six Operators III: Gauge
  Coupling Dependence and Phenomenology}},  {\em JHEP} {\bf 04} (2014) 159,
  [\href{http://arxiv.org/abs/1312.2014}{{\tt arXiv:1312.2014}}].

\bibitem{Jenkins:2017jig}
E.~E. Jenkins, A.~V. Manohar, and P.~Stoffer, {\it {Low-Energy Effective Field
  Theory below the Electroweak Scale: Operators and Matching}},  {\em JHEP}
  {\bf 03} (2018) 016, [\href{http://arxiv.org/abs/1709.04486}{{\tt
  arXiv:1709.04486}}].

\bibitem{Aebischer:2015fzz}
J.~Aebischer, A.~Crivellin, M.~Fael, and C.~Greub, {\it {Matching of gauge
  invariant dimension-six operators for $b\to s$ and $b\to c$ transitions}},
  {\em JHEP} {\bf 05} (2016) 037, [\href{http://arxiv.org/abs/1512.02830}{{\tt
  arXiv:1512.02830}}].

\bibitem{Dekens:2019ept}
W.~Dekens and P.~Stoffer, {\it {Low-energy effective field theory below the
  electroweak scale: matching at one loop}},  {\em JHEP} {\bf 10} (2019) 197,
  [\href{http://arxiv.org/abs/1908.05295}{{\tt arXiv:1908.05295}}].

\bibitem{Aebischer:2017gaw}
J.~Aebischer, M.~Fael, C.~Greub, and J.~Virto, {\it {B physics Beyond the
  Standard Model at One Loop: Complete Renormalization Group Evolution below
  the Electroweak Scale}},  {\em JHEP} {\bf 09} (2017) 158,
  [\href{http://arxiv.org/abs/1704.06639}{{\tt arXiv:1704.06639}}].

\bibitem{Jenkins:2017dyc}
E.~E. Jenkins, A.~V. Manohar, and P.~Stoffer, {\it {Low-Energy Effective Field
  Theory below the Electroweak Scale: Anomalous Dimensions}},  {\em JHEP} {\bf
  01} (2018) 084, [\href{http://arxiv.org/abs/1711.05270}{{\tt
  arXiv:1711.05270}}].

\bibitem{Aebischer:2020dsw}
J.~Aebischer, C.~Bobeth, A.~J. Buras, and J.~Kumar, {\it {SMEFT ATLAS of
  $\Delta$F = 2 transitions}},  {\em JHEP} {\bf 12} (2020) 187,
  [\href{http://arxiv.org/abs/2009.07276}{{\tt arXiv:2009.07276}}].

\bibitem{Aebischer:2021raf}
J.~Aebischer, C.~Bobeth, A.~J. Buras, J.~Kumar, and M.~Misiak, {\it {General
  non-leptonic \ensuremath{\Delta}F = 1 WET at the NLO in QCD}},  {\em JHEP}
  {\bf 11} (2021) 227, [\href{http://arxiv.org/abs/2107.10262}{{\tt
  arXiv:2107.10262}}].

\bibitem{Aebischer:2021hws}
J.~Aebischer, C.~Bobeth, A.~J. Buras, and J.~Kumar, {\it {BSM master formula
  for $\epe$ in the WET basis at NLO in QCD}},  {\em JHEP} {\bf 12} (2021) 043,
  [\href{http://arxiv.org/abs/2107.12391}{{\tt arXiv:2107.12391}}].

\bibitem{Aebischer:2022hws}
J.~Aebischer, A.~J. Buras, and J.~Kumar, {\it {NLO QCD Renormalization Group
  Evolution for Non-Leptonic $\Delta F=2$ Transitions in the SMEFT}},
  \href{http://arxiv.org/abs/2204.XXXXX}{{\tt arXiv:2204.XXXXX}}.

\bibitem{Proceedings:2019rnh}
J.~Aebischer, M.~Fael, A.~Lenz, M.~Spannowsky, and J.~Virto, eds., {\em
  {Computing Tools for the SMEFT}}, 10, 2019.

\bibitem{Aebischer:2018bkb}
J.~Aebischer, J.~Kumar, and D.~M. Straub, {\it {Wilson: a Python package for
  the running and matching of Wilson coefficients above and below the
  electroweak scale}},  {\em Eur. Phys. J.} {\bf C78} (2018), no.~12 1026,
  [\href{http://arxiv.org/abs/1804.05033}{{\tt arXiv:1804.05033}}].

\bibitem{Celis:2017hod}
A.~Celis, J.~Fuentes-Martin, A.~Vicente, and J.~Virto, {\it {DsixTools: The
  Standard Model Effective Field Theory Toolkit}},  {\em Eur. Phys. J.} {\bf
  C77} (2017), no.~6 405, [\href{http://arxiv.org/abs/1704.04504}{{\tt
  arXiv:1704.04504}}].

\bibitem{Fuentes-Martin:2020zaz}
J.~Fuentes-Martin, P.~Ruiz-Femenia, A.~Vicente, and J.~Virto, {\it {DsixTools
  2.0: The Effective Field Theory Toolkit}},  {\em Eur. Phys. J. C} {\bf 81}
  (2021), no.~2 167, [\href{http://arxiv.org/abs/2010.16341}{{\tt
  arXiv:2010.16341}}].

\bibitem{Aebischer:2017ugx}
J.~Aebischer et~al., {\it {WCxf: an exchange format for Wilson coefficients
  beyond the Standard Model}},  {\em Comput. Phys. Commun.} {\bf 232} (2018)
  71--83, [\href{http://arxiv.org/abs/1712.05298}{{\tt arXiv:1712.05298}}].

\bibitem{Buras:2020wyv}
A.~J. Buras, {\it {The $\epsilon'/\epsilon$-Story: 1976-2021}},  {\em Acta
  Phys. Polon. B} {\bf 52} (2021), no.~1 7--41,
  [\href{http://arxiv.org/abs/2101.00020}{{\tt arXiv:2101.00020}}].

\bibitem{RBC:2020kdj}
{\bf RBC, UKQCD} Collaboration, R.~Abbott et~al., {\it {Direct CP violation and
  the $\Delta I=1/2$ rule in $K\to\pi\pi$ decay from the standard model}},
  {\em Phys. Rev. D} {\bf 102} (2020), no.~5 054509,
  [\href{http://arxiv.org/abs/2004.09440}{{\tt arXiv:2004.09440}}].

\bibitem{Cirigliano:2019ani}
V.~Cirigliano, H.~Gisbert, A.~Pich, and A.~Rodr\'\i{}guez-S\'anchez, {\it
  {Theoretical status of $\varepsilon'/\varepsilon$}},  {\em J. Phys. Conf.
  Ser.} {\bf 1526} (2020) 012011, [\href{http://arxiv.org/abs/1912.04736}{{\tt
  arXiv:1912.04736}}].

\bibitem{Buras:2015xba}
A.~J. Buras and J.-M. G\'erard, {\it {Upper Bounds on
  $\varepsilon'/\varepsilon$ Parameters $B_6^{(1/2)}$ and $B_8^{(3/2)}$ from
  Large N QCD and other News}},  {\em JHEP} {\bf 12} (2015) 008,
  [\href{http://arxiv.org/abs/1507.06326}{{\tt arXiv:1507.06326}}].

\bibitem{Buras:2020pjp}
A.~J. Buras and J.-M. G\'erard, {\it {Isospin-breaking in $\varepsilon
  '/\varepsilon $: impact of $\eta _0$ at the dawn of the 2020s}},  {\em Eur.
  Phys. J. C} {\bf 80} (2020), no.~8 701,
  [\href{http://arxiv.org/abs/2005.08976}{{\tt arXiv:2005.08976}}].

\bibitem{Batley:2002gn}
{\bf NA48} Collaboration, J.~Batley et~al., {\it {A Precision measurement of
  direct CP violation in the decay of neutral kaons into two pions}},  {\em
  Phys.~Lett.} {\bf B544} (2002) 97--112,
  [\href{http://arxiv.org/abs/hep-ex/0208009}{{\tt hep-ex/0208009}}].

\bibitem{AlaviHarati:2002ye}
{\bf KTeV} Collaboration, A.~Alavi-Harati et~al., {\it {Measurements of direct
  CP violation, CPT symmetry, and other parameters in the neutral kaon
  system}},  {\em Phys.~Rev.} {\bf D67} (2003) 012005,
  [\href{http://arxiv.org/abs/hep-ex/0208007}{{\tt hep-ex/0208007}}].

\bibitem{Worcester:2009qt}
{\bf KTeV} Collaboration, E.~Worcester, {\it {The Final Measurement of
  $\varepsilon'/\varepsilon$ from KTeV}},
  \href{http://arxiv.org/abs/0909.2555}{{\tt arXiv:0909.2555}}.

\bibitem{Bona:2006ah}
{\bf UTfit} Collaboration, M.~Bona et~al., {\it {The Unitarity Triangle Fit in
  the Standard Model and Hadronic Parameters from Lattice QCD: A Reappraisal
  after the Measurements of Delta m(s) and BR(B ---\ensuremath{>} tau
  nu(tau))}},  {\em JHEP} {\bf 10} (2006) 081,
  [\href{http://arxiv.org/abs/hep-ph/0606167}{{\tt hep-ph/0606167}}].

\bibitem{Charles:2015gya}
J.~Charles et~al., {\it {Current status of the Standard Model CKM fit and
  constraints on $\Delta F=2$ New Physics}},  {\em Phys. Rev. D} {\bf 91}
  (2015), no.~7 073007, [\href{http://arxiv.org/abs/1501.05013}{{\tt
  arXiv:1501.05013}}].

\bibitem{ATLAS:2019erb}
{\bf ATLAS} Collaboration, G.~Aad et~al., {\it {Search for high-mass dilepton
  resonances using 139 fb$^{-1}$ of $pp$ collision data collected at
  $\sqrt{s}=$13 TeV with the ATLAS detector}},  {\em Phys. Lett. B} {\bf 796}
  (2019) 68--87, [\href{http://arxiv.org/abs/1903.06248}{{\tt
  arXiv:1903.06248}}].

\bibitem{CMS:2019tbu}
{\bf CMS} Collaboration, {\it {Search for a narrow resonance in high-mass
  dilepton final states in proton-proton collisions using
  140$~\mathrm{fb}^{-1}$ of data at $\sqrt{s}=13~\mathrm{TeV}$}}, .

\bibitem{Aebischer:2020lsx}
J.~Aebischer and J.~Kumar, {\it {Flavour Violating Effects of Yukawa Running in
  SMEFT}},  {\em JHEP} {\bf 09} (2020) 187,
  [\href{http://arxiv.org/abs/2005.12283}{{\tt arXiv:2005.12283}}].

\bibitem{Belle-II:2018jsg}
{\bf Belle-II} Collaboration, W.~Altmannshofer et~al., {\it {The Belle II
  Physics Book}},  {\em PTEP} {\bf 2019} (2019), no.~12 123C01,
  [\href{http://arxiv.org/abs/1808.10567}{{\tt arXiv:1808.10567}}]. [Erratum:
  PTEP 2020, 029201 (2020)].

\bibitem{CidVidal:2018eel}
X.~Cid~Vidal et~al., {\it {Beyond the Standard Model Physics at the HL-LHC and
  HE-LHC}},  \href{http://arxiv.org/abs/1812.07831}{{\tt arXiv:1812.07831}}.

\bibitem{Beacham:2019nyx}
J.~Beacham et~al., {\it {Physics Beyond Colliders at CERN: Beyond the Standard
  Model Working Group Report}},  {\em J. Phys. G} {\bf 47} (2020), no.~1
  010501, [\href{http://arxiv.org/abs/1901.09966}{{\tt arXiv:1901.09966}}].

\bibitem{USQCD:2019hyg}
{\bf USQCD} Collaboration, C.~Lehner et~al., {\it {Opportunities for Lattice
  QCD in Quark and Lepton Flavor Physics}},  {\em Eur. Phys. J. A} {\bf 55}
  (2019), no.~11 195, [\href{http://arxiv.org/abs/1904.09479}{{\tt
  arXiv:1904.09479}}].

\bibitem{Cirigliano:2019jig}
{\bf USQCD} Collaboration, V.~Cirigliano, Z.~Davoudi, T.~Bhattacharya,
  T.~Izubuchi, P.~E. Shanahan, S.~Syritsyn, and M.~L. Wagman, {\it {The Role of
  Lattice QCD in Searches for Violations of Fundamental Symmetries and Signals
  for New Physics}},  {\em Eur. Phys. J. A} {\bf 55} (2019), no.~11 197,
  [\href{http://arxiv.org/abs/1904.09704}{{\tt arXiv:1904.09704}}].

\bibitem{Joo:2019byq}
{\bf USQCD} Collaboration, B.~Jo{\'o}, C.~Jung, N.~H. Christ, W.~Detmold,
  R.~Edwards, M.~Savage, and P.~Shanahan, {\it {Status and Future Perspectives
  for Lattice Gauge Theory Calculations to the Exascale and Beyond}},  {\em
  Eur. Phys. J.} {\bf A55} (2019), no.~11 199,
  [\href{http://arxiv.org/abs/1904.09725}{{\tt arXiv:1904.09725}}].

\end{thebibliography}\endgroup

\end{document}